\documentclass[fleqn, twocolumn, a4paper, 10pt]{wlscirep}
\usepackage{setspace}
\usepackage[utf8]{inputenc}
\usepackage[T1]{fontenc}
\usepackage{ amssymb }
\usepackage{siunitx}
\usepackage{amsmath}
\usepackage{upgreek}
\usepackage{graphicx,subcaption,lipsum}
\usepackage{threeparttable}
\usepackage{color} 
\usepackage{soul} 
\usepackage{ulem}

\usepackage{cuted}

\title{Symmetry breaking in Prussian Blue Analogues via growth--guided local ordering of hexacyanometallate vacancies
}
\usepackage{longtable}
\author[1]{Yevheniia Kholina}
\author[1]{Thomas Weber}
\author[1]{Joohee Bang}
\author[2]{Arthur Baroni}
\author[2,3]{Marianne Liebi}
\author[4]{Semen Gorfman}
\author[4]{Ido Biran}
\author[5]{Mark Warren}
\author[6]{Dmitriy Chernyshov}
\author[1,*]{Arkadiy Simonov}
\affil[1]{ETH Z{\"u}rich, Department of Materials, Vladimir-Prelog-Weg 4, Z{\"u}rich, 8093, Switzerland}

\affil[2]{Photon Science Division, Paul Scherrer Institut, 5232 Villigen PSI, Switzerland}
\affil[3]{Institute of Materials, Ecole Polytechnique Fédérale de Lausanne (EPFL), 1015 Lausanne, Switzerland}

\affil[4]{Tel Aviv University, Department of Materials Science and Engineering, Ramat Aviv, Tel Aviv, 6997801, Israel}
\affil[5]{Diamond Light Source, Harwell Campus, Chilton, Oxfordshire OX11 0DE}
\affil[6]{European Synchrotron Radiation Facility, Grenoble, 38043, France}

\affil[*]{arkadiy.simonov@mat.ethz.ch}

\begin{abstract}
We report Growth--Guided Local Ordering, a novel mechanism of symmetry reduction in disordered crystals. This mechanism operates through the directional ordering of point defects during crystal growth, where defect correlations develop preferentially along the growth direction, resulting in reduced symmetry that persists in the final structure through the spatial distribution of defects.

We demonstrate this phenomenon in Mn[Co]-Prussian Blue Analogues, disordered cyanide crystals containing numerous Co(CN)$_6$ vacancies. Single crystal diffuse scattering reveals pronounced anisotropy in vacancy distribution: strong correlations along [001] growth direction contrast with weak correlations perpendicular to it. This local ordering reduces the Laue symmetry to tetragonal $4/mmm$, evident in properties such as birefringence, while the average structure retains cubic $m\bar 3m$ symmetry. When growth proceeds along [111] direction, the same mechanism produces domains with trigonal symmetry. Because this mechanism relies on fundamental aspects of crystal growth rather than specific material properties, it offers a general strategy for symmetry control in disordered crystals. Crucially, it transforms the complex task of altering crystal symmetry into the more manageable challenge of controlling growth direction, achievable through various established techniques such as the use of surfactants during crystallization.

\end{abstract}
\begin{document}

\flushbottom
\maketitle

\thispagestyle{empty}

\section*{Introduction}
While symmetry is beautiful, breaking of symmetry often leads to the emergence of novel functional properties. This is fundamental across scales---from materials exhibiting birefringence \cite{nye1985physical}, magnetism, and ferroelectricity \cite{fiebig2023nonlinear}, to phenomena in spintronics \cite{bloom2024chiral}, superconductivity \cite{vojta2009lattice}, and catalysis \cite{mikami2003symmetry}, extending even to the emergence of mass in elementary particles \cite{nambu1961dynamical, higgs1964broken, livio2012symmetry}.

In the synthesis of bulk crystals, symmetry reduction remains challenging as it typically relies on spontaneous symmetry breaking mechanisms, such as octahedral tilts in perovskites or cooperative Jahn-Teller distortions \cite{goodenough1998jahn, benedek2022hybrid, bostrom2020tilts, bostrom2018recipes}, which are inherently difficult to control. By the very nature of structural phase transitions, a new symmetry is set by the long-range ordering of local distortions. Here we propose an alternative approach that can be employed in disordered crystals, where symmetry reduction is achieved through growth--guide short-range ordering of defects.

\begin{figure*}[]
\centering
\includegraphics[width=0.95\linewidth]{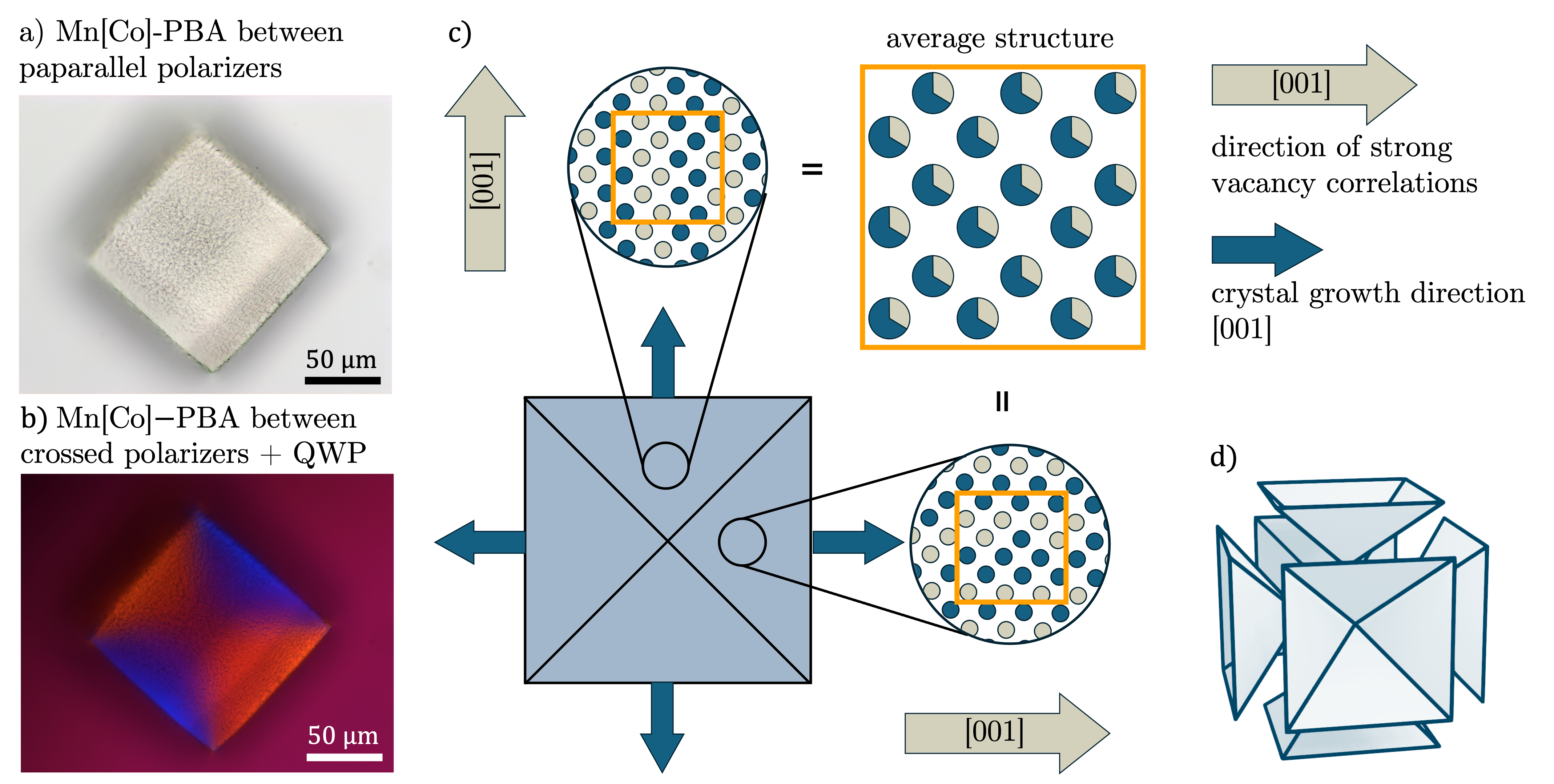}
\caption{Mn[Co]-PBA crystals grow in the shape of cubes (a). When placed between crossed linear polarizers, these crystals show weak optical birefringence and two-color contrast when in addition to crossed polarizers a quarter-wave plate (QWP) is inserted (b). Such optical contrast suggests that each crystal cube consists of six square-pyramidal domains (c,d). Using Bragg and diffuse scattering analysis we show that all six domains share the same average structure with the cubic spacegroup $Fm\bar 3m$. However, their optical properties are dictated by the local structure which is tetragonal. This symmetry lowering is caused by the anisotropic ordering of hexacyanometallate vacancies with strong correlation along the [001] growth directions (c).}
\label{fig:birefr}
\end{figure*}

We demonstrate this method on the ordering of hexacyanometallate vacancies in a disordered Prussian Blue Analogue (PBA). We demonstrate that despite the average cubic structure, the local structure is highly anisotropic. The vacancies order preferentially along the crystal growth direction, resulting in a symmetry reduction from cubic to tetragonal (Figure \ref{fig:birefr}). This symmetry reduction manifests itself in optical properties such as the appearance of birefringence highlighting the potential of Growth--Guided Local Ordering for tailoring material properties.

Prussian Blue Analogues are a wide family of cyanide materials primarily investigated for their potential applications in gas storage \cite{chapman2005reversible}, gas purification \cite{motkuri2010dehydrated}, as catalysts \cite{Goberna-Ferron2014, Guo2020}, and as electrode materials \cite{cattermull2021structural, cattermull2023k}. Some PBA representatives also exhibit rare magnetic and electronic effects like charge transfer\cite{sato1996photoinduced, aguila2016switchable} or humidity-sensitive magnetism\cite{Stefanczyk2019}. 

Extensive research using conventional Bragg diffraction has revealed that PBAs can adopt cubic, orthorhombic, or rhombohedral symmetries \cite{cattermull2021structural}. In PBA battery cathode materials, the symmetry is often reduced to orthorhombic or rhombohedral due to cooperative A-site ions displacements \cite{cattermull2021structural}, octahedral tilts, cooperative Jahn–Teller distortions, columnar shift, etc. However, in PBAs containing large number of vacancies, these vacancies are typically considered disordered and the symmetry is reported to be $Fm\bar{3}m$. The model system in the current study is Mn[Co] Prussian Blue Analogue with a composition Mn[Co(CN)$_6$]$_{2/3} \cdot n$H$_2$O belongs to the latter and is well-known to have a cubic average structure in which Co and Mn atoms occupy alternating sites in a simple NaCl-type arrangement and are connected by cyanide linkers, as shown in Figure \ref{fig:av_str}.

The rest of the paper is organized as follows. We first use optical measurements to reveal unexpected birefringence in cubic PBA crystals, demonstrating deviation from cubic symmetry. Through combined Bragg and diffuse X-ray scattering analysis, we show that while the average structure remains cubic, the local structure is tetragonal due to preferential ordering of Co(CN)$_6$ vacancies along the [001] growth direction. Finally, we demonstrate that by growing crystals with \{111\} faces, we can achieve domains with trigonal symmetry, establishing growth direction as a powerful parameter for controlling crystal symmetry.

\section*{Main}
\subsection*{Optical Observations}

We have grown single crystals of Mn[Co] Prussian Blue Analogue (Mn[Co]-PBA) in Agar gel media, resulting in well-formed cubes (Figure \ref{fig:birefr}a). When placed between crossed polarizers, these cubic crystals show a distinct light contrast pattern resembling pyramids (Figure \ref{fig:birefr}b). This optical property, known as birefringence, emerges only in non-isotropic media. Birefringence implies an anisotropic direction in the crystal and indicates that the real symmetry of the given crystals is lower than the previously assumed cubic symmetry.

Mn[Co]-PBA crystals in crossed polarizers show a four-triangle contrast pattern, independent of which side the crystal is placed on. Such a contrast in cube-shaped crystals suggests that each cube consists of six square-based pyramid-shaped twins, likely having tetragonal crystal symmetry (Figure \ref{fig:birefr}c,d).

This observation apparently contradicts the well-known fact that disordered PBAs have $Fm\bar{3}m$ symmetry in their average structure. To reconcile this discrepancy, we first employed classical diffraction approaches.

\subsection*{Average Structure}
\subsubsection*{High-resolution single-crystal XRD}

To investigate potential deviations of Mn[Co]-PBA from cubic symmetry, we performed high-resolution reciprocal space mapping around 18 different Bragg peaks. Any deviation from cubic symmetry would manifest through the formation of strain or ferroelastic domains, resulting in the corresponding splitting of the Bragg peaks.

In all cases, the observed Bragg peaks were distributed over several detector pixels without any apparent peak splitting. We analyzed the corresponding rocking curves, which represent the $\omega$-dependencies of diffraction intensities integrated over the relevant detector area. Supplementary Figure \ref{fig:peak_splitting} illustrates three such rocking curves from the 222, 026, and 080 Bragg peaks, corresponding to scattering angles of 30°, 55°, and 72°, respectively. The width of each rocking curve is below 0.02°, which is close to the instrument's resolution limit. For comparison, we included the rocking curves of the 111, 113, and 133 reflections from a silicon crystal, measured at similar scattering angles.

These high-resolution measurements demonstrate the absence of peak splitting, strongly indicating the absence of strain domains and supporting the preservation of cubic symmetry in the average structure of Mn[Co]-PBA (Figure \ref{fig:av_str}).

\subsubsection*{Single-crystal Bragg diffraction from untwinned portion}

To further investigate the apparent discrepancy between optical observations and the average structure, we performed single-crystal X-ray diffraction (XRD) on an isolated section of the crystal. This section, approximately 40 × 50 × 40 $\mu$m in size, was carefully cut from a bigger crystal to correspond to only one of the six pyramidal domains observed optically. The measured dataset showed reliability factors R$_{int}$(cubic) = 3.28\% and R$_{int}$(tetragonal) = 3.16\%, indicating negligible preference for tetragonal symmetry. The Bragg peak intensity distribution followed cubic symmetry, and automatic Laue group analysis resulted in the $m\bar 3m$ point group, consistent with cubic symmetry. 

Structure refinement in the cubic space group $Fm\bar 3m$ yielded wR2(cubic) = 12.87\% (tables \ref{table:str_cub} and \ref{table:str_cub_tetr} with the refined structure parameters in cubic and tetragonal settings respectively). After transforming the refined structure to tetragonal $I4/mmm$ group and performing refinement with this lower symmetry, we obtained a margnianlly better wR2(tetragonal) = 12.76\%. The tetragonal structure revealed no significant displacements of the atomic positions (table \ref{table:str_tetr} and Figure \ref{fig:cubic_vs_tetragonal}). The R-factor improvement (table \ref{table:refinem_params}, Figure \ref{fig:fobs_calc}) came from a slight change in anisotropic displacement parameters (ADPs) attributable to systematic errors of the experiment. 

In short, conventional diffraction approaches do not detect any clear indications of symmetry breaking from cubic to tetragonal in the average structure, despite the optical evidence suggesting lower symmetry.

\subsection*{Local Structure}

To resolve the apparent contradiction between optical properties and the symmetry suggested by diffraction experiments, we measured diffuse scattering as a probe of local structure from the untwinned portion of the crystal.

The Co(CN)$_6$ vacancies, though disordered on average, have a strong local order \cite{simonov2020hidden}. In diffraction experiments, this local order manifests itself as characteristic "squares" of diffuse scattering around the Bragg peaks (Figure~\ref{fig:diffuse}a). Unlike the Bragg peaks, which follow cubic symmetry, the diffuse scattering clearly shows tetragonal symmetry. Notably, the pattern is sharp along the $\mathbf c$-axis --- which corresponds to the crystal's growth direction --- but broad in other directions.

This diffuse signal can be explained by correlations between Co(CN)$_6$ occupancies and indicates their non-random distribution in the network on the local scale. Anisotropy of diffuse scattering shape reflects a highly ordered arrangement of hexacyanometallate ions and vacancies along the growth direction, contrasting with weaker correlations along other axes Figure~\ref{fig:birefr}c). This alignment pattern likely reflects the crystal assembly process. During growth, the incomplete surface layer necessarily exhibits lower (tetragonal) symmetry. While most crystals develop full cubic symmetry as subsequent layers form, here the initial surface-imposed arrangement of defects appears to become locked in, preserving a memory of the growth surface symmetry within the cubic average structure.

\subsubsection*{3D-$\Delta$PDF Analysis}

To quantify the degree of local order, we employed three-dimensional difference pair distribution function analysis (3D-$\Delta$PDF). This method provides information about the spatial correlations in the distribution of Co(CN)$_6$ clusters and their vacancies (Figure \ref{fig:MC_simulated_crossections}b), as well as correlations in static displacements around these vacancies. To obtain this 3D-$\Delta$PDF map, we first refined all substitutional and displacive correlations simultaneously for all the pairs available in the direct space (Figures \ref{fig:diffuse_modelled_PDF}). Afterwards we separate the signal into two portions, the diffuse signal arising due to the substitutions correlations and the displacive correlations (Figure \ref{fig:models_separated}). The analysis of displacive correlations is not covered in this work and can be found in \cite{kholina2025dehydration}.

\begin{figure*}[!ht]
\centering

\includegraphics[width=0.95\linewidth]{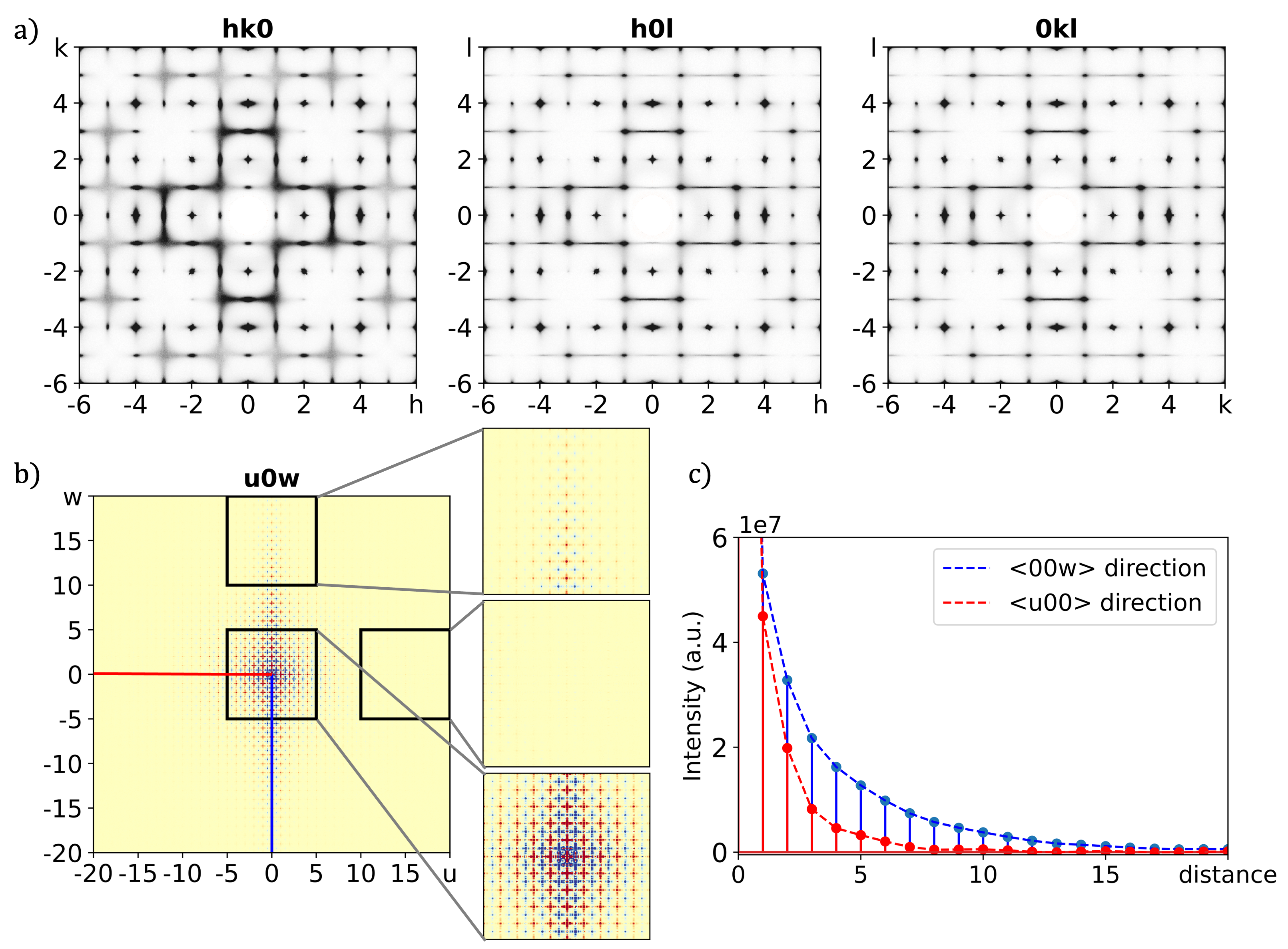}

 \caption{Single crystal diffuse scattering from Mn[Co]-PBA. a) Three crosssections, hk0, h0l, and 0kl of the reconstructed three-dimensional diffuse scattering dataset. This dataset contains Bragg peaks at even h,k,l, and diffuse scattering located in between. The characteristic diffuse scattering on a border of the Brillouin zone can be explained by non-random Co(CN)$_6$ distribution. The intensity distribution of this diffuse scattering follows tetragonal symmetry, indicating anisotropic vacancy distribution in Mn[Co]-PBA. The difference in diffuse scattering symmetry can be well seen by comparing the diffuse feature around (h+1, k+1, 0, with h,k = odd) on hk0 plane versus h0l plane. Diffuse scattering becomes sharp along l direction, which breaks 3-fold rotation along <111> directions. We demonstrate the difference in the length of vacancy correlations along two perpendicular directions (c) by plotting the profiles of 3D-$\Delta$PDF signals which indicate the strength of vacancy correlations, color code is the same as in (b).}
\label{fig:diffuse}
\end{figure*}

The 3D-$\Delta$PDF map, Figure \ref{fig:diffuse}b, reveals a non-random distribution of vacancies characterized by short-range vacancy-vacancy avoidance. This ordering pattern is most pronounced for the first neighboring distance, corresponding to interatomic vectors such as 1/2 0 1/2. Specifically, if a certain site is vacant, its nearest neighbors are more likely to be occupied than it would be expected from a random distribution of vacancies.

For second-nearest neighbors, we observe positive correlations, with slightly stronger correlations along the [001] direction compared to the [100] and [010] directions. Beyond this range, the decay of correlations exhibits significant anisotropy: along the [001] direction, correlations persist to a correlation length of approximately 18 unit cells, while in the [100] and [010] directions, the correlation length is only about 8 unit cells (Figure \ref{fig:diffuse}c).

table \ref{table:pair_correlations}

This anisotropy in vacancy ordering induces statistically tetragonal $4/mmm$ Laue symmetry in the crystal at the local scale, which manifests macroscopically as birefringence. Furthermore, this directional ordering has implications for the pore network characteristics: vacancy channels exhibit slightly better connectivity within planes perpendicular to the $\mathbf c$-axis than along the $\mathbf c$-axis itself.

\subsection*{Real-Space Model}

To obtain a real-space representation of the local ordering, we performed Monte Carlo (MC) simulations based on a model similar to that described in our previous work by Simonov et al. \cite{simonov2020hidden}. The MC model incorporates two terms:
\begin{itemize}
    \item J$_1$: A first-nearest neighbor interaction enforcing electrostatic neutrality and driving the main feature of local order - negative correlations between nearest-neighbor vacancies.
    \item J$_2$: A weak next-nearest neighbor interaction describing positive vacancy correlations along cubic <100> directions. To account for the observed tetragonal symmetry, we split this parameter into $J_{2,x}=J_{2,y}$ and $J_{2,z}$.
\end{itemize}

To reproduce the observed diffuse scattering pattern, we found that the strength of $J_{2,z}$ must be set approximately six times of $J_{2,xy}$. 

\begin{figure*}[!ht]
\centering
\includegraphics[width=\linewidth]{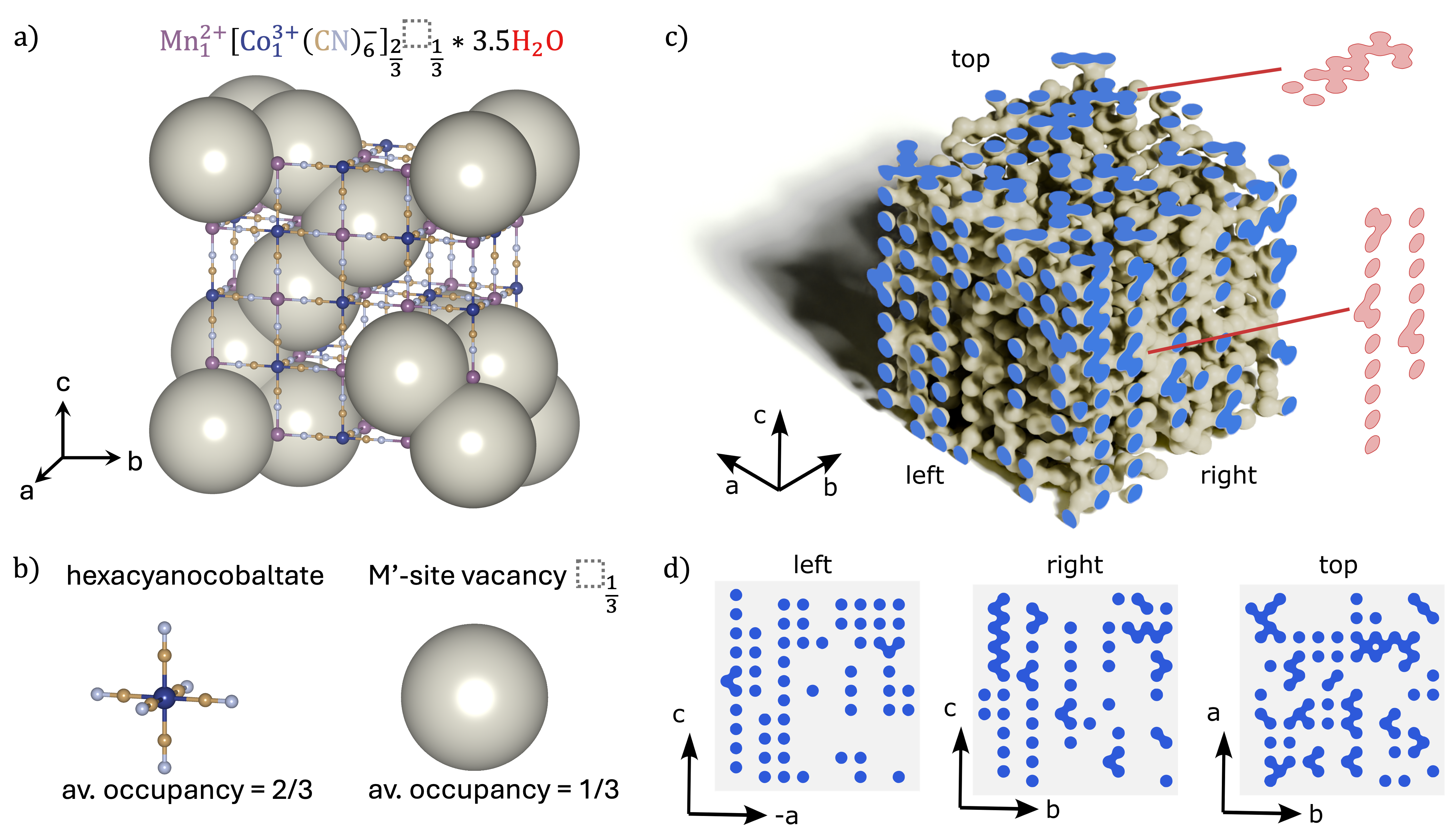}
\caption{(a) The Mn[Co]-PBA structure consists of Mn and Co ions linked by cyanides. (b) Due to electroneutrality condition, one-third of hexacyanocobaltate clusters are absent and replaced with large vacancies which we represent as beige spheres. (c) Monte Carlo simulations of a supercell of Mn[Co]-PBA show highly anisotropic distribution of the vacancies. For clarity we remove all atoms and only show the vacancies, moreover three vacancy planes which are facing the viewer are colored in blue. In red we present characteristic motifs of the vacancy distributions: long chains of ordered vacancies along the $\mathbf z$ axis evident on the "left" and "right" faces, and symmetric disrtibution on the top face. (d) The same pattern can be seen in the cross-sections of the model: on the left and right face, the predominant motif is one dimensionally ordered chains of vacancies stretched along the $\mathbf z$ axis, while on the top face, the vacancies are distributed symmetrically. Note that we use the term "vacancy" to indicate an empty hexacyanocobaltate site. In reality, these vacancies are filled with water molecules.}
\label{fig:MC_simulated_crossections}
\end{figure*}

Figure \ref{fig:MC_simulated_crossections} illustrates the resulting real-space vacancy configurations in three perpendicular cross-sections. In the $xy$ plane, vacancies exhibit no preferential ordering direction but tend to avoid one another in a checkerboard-like pattern, resulting in diagonally connected pore channels. In contrast, the $xz$ and $yz$ cross-sections reveal strong vacancy ordering along the [001] direction, with the enhanced $J_{2,z}$ interaction ensuring weaker vacancy connectivity along this axis.

Figure \ref{fig:MC} compares the calculated diffuse scattering from these simulated structures with experimentally measured diffuse scattering, demonstrating a good agreement.

The [001] direction, characterized by longer vacancy correlations, corresponds to the crystal growth direction. This suggests that during crystal formation, incoming Co(CN)$_6$ building blocks are more strongly influenced by the already-incorporated blocks beneath them than by neighboring blocks on the surface. Such behavior is consistent with a layer-by-layer growth mode.

As the crystal grows simultaneously in all six directions, the resulting cube comprises six square-based pyramids. Each pyramid has a tetragonal direction with a high degree of vacancy order pointing outward from the cube face. This represents a highly unusual type of twinning: while traditional twins comprise domains with the same low-symmetry average structure in different orientations, here all domains share an identical cubic average structure but differ in the orientation of their local vacancy ordering. The twin boundaries at cube edges form more defective regions where these different ordering patterns meet, explaining the previously observed selective etching along cube edges in Fe[Fe]-PBA microcrystals \cite{hu2010prussian}.

To our knowledge, examples of such symmetry breaking due to local ordering in crystals are extremely rare. This scarcity is particularly striking given that computational growth models of disordered crystals consistently predict symmetry reduction along the growth direction \cite{welberry2022diffuse}. Theory suggests that, unless the model is specifically constructed \cite{welberry19793}, the symmetry of disordered crystals must always reflect the growth process. Yet experimental observations of this effect have remained elusive. Hulliger et al. described similar effects in host-guest organic molecular crystals \cite{hulliger2000polarity, rechsteiner2000phase, hulliger1997crystallization, roth1998statistically}. However, in their work, the low symmetry arises from the alignment of asymmetric molecules withing the host network, directly affecting the average structure. Our work demonstrates symmetry lowering in a crystal with a cubic lattice and highly symmetric cubic building blocks. The symmetry of these crystals is defined solely by the arrangement of Co(CN)$_6$ building blocks during the crystal assembly process.

We propose that this apparent scarcity of experimental observations may be due to the inherent difficulty in detecting these effects: since these crystals invariably grow as twins, their structural anisotropy is effectively masked in average structure measurements. The symmetry reduction only becomes apparent when examining diffuse scattering from individual twin components, as we have done here. This suggests that growth--guided symmetry breaking may be far more prevalent than previously recognized, but has simply gone undetected due to the limitations of conventional characterization approaches.

\subsection*{Symmetry Control Through Growth Direction}

So far we have shown how growth along the $[100]$ direction results in crystals with tetragonal symmetry, but we can use it to achieve different symmetries by selecting another growth direction. When grown with excess MnCl$_2$, Mn[Co] PBAs develop into cubeoctahedra with both \{111\} and \{100\} faces. The relative size of the \{111\} faces can be controlled through the Mn$^{2+}$:Co(CN)$_6^{3-}$ concentration ratio.

\begin{figure*}[!ht]
\centering
\includegraphics[width=\linewidth]{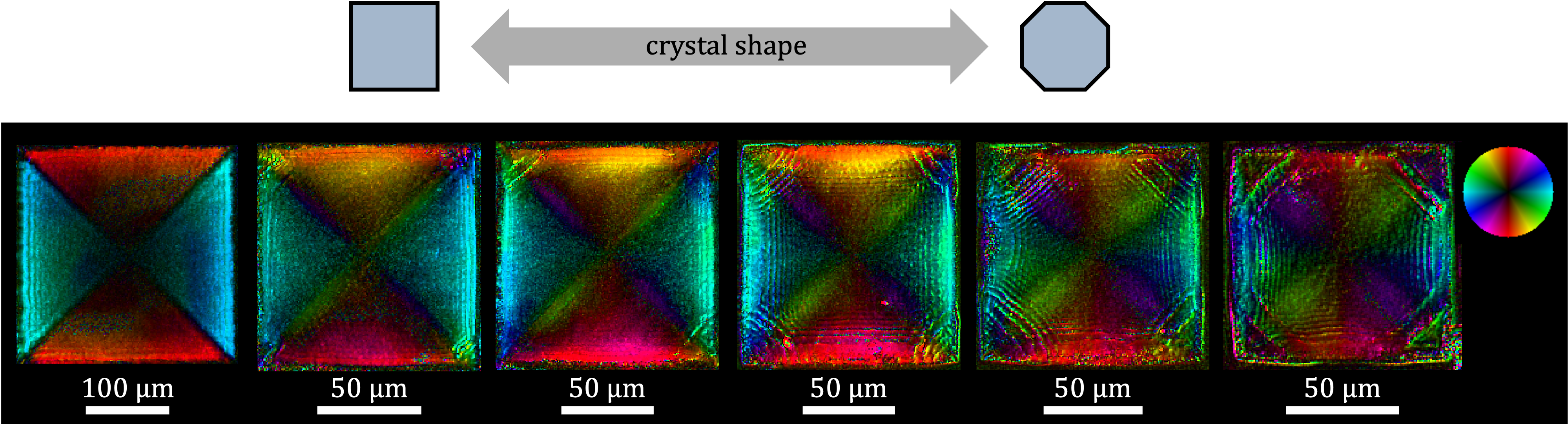}
\caption{Mueller polarimetry of Mn[Co]-PBA crystals reveals the relationship between crystal morphology and domain structure. The color wheel (top right) indicates the direction of the fast optical axis, while color saturation represents retardance. Cubic crystals (left) exhibit four clearly visible triangular domains, with light blue domains indicating left-right fast axis orientation and red domains showing top-bottom orientation. Together with two domains along the viewing direction, this forms a pattern of six square-pyramidal domains, each with its fast axis aligned along the respective growth direction. Crystals with increasing degrees of truncation develop additional diagonal domains. These appear as thin lines at domain boundaries, colored bright pink and green, indicating fast axes oriented perpendicular to cube diagonals. The size of these diagonal domains increases with the degree of crystal truncation, proportional to the size of {111} faces.}
\label{fig:Mueller_microscopy}
\end{figure*}

Mueller polarimetry microscopy provides evidence for new symmetry domains in these cubeoctahedral crystals (Figure \ref{fig:Mueller_microscopy}). The false colors in the crystal images indicate the direction of the fast optical axis, i.e., the orientation where polarized light experiences the lowest refractive index and thus travels fastest through the material, and the intensity of the color is proportional to the value of retardance (the brighter colors relate to less retardant, black - to non-retardant areas). In each domain of the cube-shaped crystal, the fast axis points along the growth direction, toward the nearest \{100\} face. In crystals with small \{111\} faces (Figure \ref{fig:Mueller_microscopy}), in addition to four \{100\}-type domains, one can see additional thin domains stretching along the cube diagonals. These new domains correspond to growth toward the \{111\} faces. In crystals with larger \{111\} faces, these diagonal domains become more pronounced, and the \{100\} faces show a more complex color gradient that likely reflects the interaction between signals from different domain types through the crystal depth (Figure \ref{fig:Mueller_microscopy}). Interestingly, Mueller matrix analysis shows that in contrast to \{100\}-type domains, in which the fast axis was pointing towards the growth direction, in \{111\}-type domains their \textit{slow} axis is oriented along to the growth direction. 

Symmetry analysis shows that $\bar3m$ is the highest symmetry subgroup of the cubic structure that could describe \{111\} domains, a prediction strongly supported by our optical measurements. This indicates that growth along $<111>$ directions likely produces domains with trigonal symmetry, analogous to how $<100>$ growth yields tetragonal domains. Unfortunately, the crystals exhibiting these \{111\} domains are quite small and possess complex overlapping domain patterns. These limitations currently prevent us from performing detailed diffuse scattering analysis that would reveal the three-dimensional distribution of vacancies within these domains. However, future diffuse scattering tomography studies would be valuable to map out these vacancy distributions and fully understand how growth direction controls the symmetry of the resulting domains.

These findings demonstrate that growth direction serves as a powerful parameter to engineer crystal symmetry. This represents an entirely new approach to symmetry control - one that operates through the organization of defects during crystal growth. The method is remarkably practical, as growth direction for bulk crystals can be controlled through standard crystallization techniques, like surfactant addition; for epitaxial thin films the growth direction can be controlled by a substrate. This provides a straightforward yet powerful way to tailor symmetry-dependent properties in disordered materials.

\section*{Discussion and outlook}
Crystal symmetry is conventionally understood through the lens of average structure, but local structural correlations can independently drive symmetry reduction. Here we demonstrate how the spatial distribution of point defects can become a powerful tool for symmetry control, even in materials where the average structure maintains high symmetry. We demonstrate this effect in Mn[Co]-Prussian Blue Analogues. While classical crystallographic approaches indicate that the average structure is cubic with space group $Fm\bar{3}m$, our diffuse scattering analysis reveals a different picture at the local scale. Vacancies exhibit strong alignment along the crystal growth direction and shorter-range correlations in the perpendicular planes. In cubic crystals this anisotropic vacancy distribution effectively reduces the local symmetry to tetragonal $4/mmm$ and results in enhanced pore connectivity within the planes perpendicular to the growth direction. In cubeoctahedral crystals, alongside tetragonal domains, this leads to the formation of domains with trigonal symmetry.

This observation exemplifies a broader effect we term 'Growth--Guided Local Ordering' which can control symmetry in disordered crystals. The effect relies on two key conditions: first, the crystal must exhibit a disordered average structure; second, the growth process must lock in the arrangement of disordered atoms or molecules, preventing subsequent reorganization. When the disordered building blocks are highly symmetric, as in Mn[Co]-PBA, this mechanism leads to symmetry reduction solely in the local structure while preserving the high symmetry of the average structure. In our case, the insoluble nature of the material ensures that once the building blocks attach during growth, they do not rearrange, thus preserving the growth-induced anisotropy.

The straightforward requirements suggest broad applicability of the principle across various material systems, particularly those grown out of equilibrium where post-growth annealing (which would allow rearrangement) is avoided. In this paper, we demonstrated it for the crystal grown from solution, but we expect it to also work for electrochemical growth and growth from flux. Crucially, it transforms the complex task of altering crystal symmetry into the more manageable challenge of controlling growth direction, achievable through various established techniques such as the use of surfactants during crystallization, and thus provides an elegant solution to the complex task of engineering local symmetry in disordered crystals.

\section*{Methods}

\subsection*{Synthesis}
The single crystals of Mn[Co]-PBA were grown by the slow counter-diffusion of aqueous solutions of manganese (II) chloride, MnCl$_2$ (Fluka), and potassium hexacyanocobaltate (III), K$_3$[Co(CN)$_6$] (Sigma-Aldrich), in Agar gel media at 2$^\circ$C. All chemicals were used without further purification. The growth was performed in 15 ml centrifuge tubes: first, the MnCl$_2$ (500\,$\upmu$mol) was dissolved in 1\% Agar gel (10\,ml) and poured into the centrifuge tube. After the gel was set, the water solution of Potassium K$_3$[Co(CN)$_6$] (500\,$\upmu$mol in 5\,ml H$_2$O) was added on top. The salts diffused through the gel, resulting in precipitation of insoluble transparent cubic crystals in a course of two weeks. The crystals used for the measurements have a side length of around 100-150\,$\upmu$m and were extracted from the lower part of the centrifuge tube.

\subsection*{Single crystal XRD and Single crystal diffuse scattering measurements}
Preliminary diffraction measurements of the Mn[Co]-PBA crystals were performed at the Swiss-Norwegian beamline (BM01) \cite{dyadkin2016new} at the European Synchrotron Radiation Facility (ESRF), 
while the final datasets were collected at the beamline I19, at the Diamond light source. The datasets were measured at a wavelength $\lambda$ = 0.71 \AA \ and beamsize 260 x 73 $\upmu$m with Eiger2 4M CdTe detector. Bragg and diffuse scattering datasets were collected with full and attenuated beam (1$\%$) respectively. Each collection consisted of a single 360$^\circ$ rotation scan around the $\phi$ axis with a 0.1$^\circ$ step per frame. 

Single crystals of 150 $\upmu$m were cut with a scalpel under a polarizing microscope to get the untwined portion of a crystal. The crystals were glued to the Mitegen loops \cite{MiTeGen} using two component epoxy glue.

Crystal orientation, indexing, and integration of the Bragg peaks were performed using the program XDS \cite{kabsch2010xds} and Dials \cite{winter2018dials}; the reconstruction of diffuse scattering reciprocal space maps was carried out using the program Meerkat \cite{simonov-meerkat}.

\subsection*{Average structure analysis}

Structure solution and refinement of Mn[Co]-PBA were done using the SHELX software suite \cite{sheldrick2008short} in conjunction with the Olex2 graphical interface \cite{dolomanov2009olex2}. Using the same Bragg reflections, we performed refinements using cubic $Fm \bar 3m$ and tetragonal $I4/mmm$ space groups. We took special care to ensure that the the scaling in XDS \cite{kabsch2010xds} was made in tetragonal symmetry and tetragonal axis is pointing along the unique $00l$ direction (which can be seen from the reconstructed hk0 plane in figure \ref{fig:projection}).

\subsection*{High-resolution single-crystal X-ray diffraction}
High-resolution single-crystal X-ray diffraction measurements were conducted using a custom-built four-circle X-ray diffractometer at Tel Aviv University \cite{gorfman2021multipurpose} using a PILATUS 1M pixel area detector. The diffractometer is equipped with a double crystal monochromator, and was specifically designed to combine standard single-crystal X-ray diffraction measurements with high-resolution reciprocal space mapping capabilities. Various benchmarks \cite{gorfman2022identification, biran2024lattice, gorfman2021multipurpose} demonstrate that the horizontal and vertical divergence of the beam are approximately 0.01$^\circ$, with a relative wavelength dispersion of $2 \cdot 10^{-4}$.

A crystal of approximately 100 $\mu$m was attached to a double-sided tape and a flat-based metal holder, which provided better stability compared to the more standard MitiGen loops. The measurements began with the determination of the crystal orientation matrix, followed by high-resolution scans around specific Bragg peaks of interest. These peaks were chosen to represent a wide range of scattering angles (16$^\circ$--99$^\circ$) and various combinations of the reflection $hkl$ indices.

\subsection*{Three-dimensional pair distribution function analysis}
Correlated disorder in Mn[Co]-PBA crystals was analyzed using Yell software \cite{simonov2014yell}, which relies on the Three Dimensional Difference Pair Distribution function (3D-$\Delta$PDF) approach \cite{weber2012three}.
3D-$\Delta$PDF is a Fourier transform of diffuse scattering and can be understood as the difference between the Pair Distribution Function of the real crystal structure and the Patterson function calculated from its average structure. 3D-$\Delta$PDF allows local structure analysis and refinement expressed in pair probabilities of two atoms being found at the given lattice vectors. Each signal in the 3D-$\Delta$PDF at a position $uvw$ corresponds to a pair of atoms $ij$ in the structure such that $x_j-x_i=u$, $y_j-y_i=v$ and $z_j-z_i=w$. Signals in the 3D-$\Delta$PDF can be positive or negative: for positive (negative) signals the interatomic pairs appear more (less) frequently in the real structure than in the average structure. For a more in-depth introduction see \cite{weber2012three}. 

For Figure 2 the  3D-$\Delta$PDF was calculated by punching out the Bragg peaks alongside the broad diffuse around Bragg peaks with large
spheres and applying Fourier transform to get a three-dimensional map of interatomic lattice vectors with a slowly decaying signal, which
indicates the strength of vacancy correlations. The reconstruction was done on the fine grid (801x801x801 pixels) to get access to long interatomic vectors.

For 3D-$\Delta$PDF quantitative refinement, we have prepared a coarser reconstruction of diffuse scattering. We cleaned it from Bragg peaks by subtracting the area of 3 voxels centered at each Bragg peak by punch-and-fill procedure, and subtracted the spherically symmetric background.  We used the refined average structure from Bragg diffraction as the basis for the model. We refined all substitutional and displacive correlations simultaneously (Figure \ref{fig:diffuse_modelled_PDF}) for all the pairs available in the direct space. The separated portions of the diffuse signal arising due to the substitutions correlations and the displacive correlations are represented in Figure \ref{fig:models_separated}. The analysis of displacive correlations is not covered in this work and can be found in \cite{kholina2025dehydration}.

\subsection*{Monte Carlo modelling}
The models of vacancy distribution for Mn[Co]-PBA were generated using Monte Carlo modeling. The custom scripts are included in supplementary files. The pair interaction parameters for this Monte Carlo were calculated using the mean field approximation. The details about this procedure are provided elsewhere \cite{kholina2025dehydration}.

The model is represented by an Ising-like Hamiltonian with pair-wise interactions. We include nearest neighbors and second nearest neighbors interactions: $J_{1}=0.605$, $J_{2,x}=J_{2,y}=-0.0135$, and $J_{2,z}=-0.11$, while temperature $T$ is set to 1.3. We generate 5,000 structures for the given parameters, each structure is represented by a 16x16x16 supercell. Configurations are initialized with a random distribution of the vacancies such that one-third of the sites are vacant. Monte Carlo steps involve swap moves: one occupied and one vacant are selected at random, and their contents are swapped. The configurations were equilibrated for 500 epochs (one epoch corresponds to the number of steps required to visit each site twice on average) with 20 epochs steps in between each produced structure. 

The model of the structure is approximate, the aim is not to represent an optimal model, but the model that captures the essence of the ordered structure.
The comparison between experimental diffuse scattering and the diffuse scattering calculated from the structures generated by Monte Carlo is shown in Figure \ref{fig:MC}.

\subsection*{Mueller polarimetry microscopy}
The polarimetric measurements were conducted with a custom-built Mueller polarimeter, based on the inverted microscope Olympus IX73 (Figure \ref{fig:MP_setup}). The probing of the sample was performed with an incoherent narrow bandwidth red light source (non-polarized diode at 625nm, Thorlabs MCS103), coupled with an optical fibre (Thorlabs M122L02) and collimated with a collimator (Thorlabs PAF2P-A15A), tilted to normal incidence toward the sample with a dielectric mirror at 45$^o$ (Thorlabs CCM1-E02/M). The polarization of the probe, referred to as polarisation state generated (PSG), is controlled by a linear polarizer, PSG LP (Newport - 10LP-VIS-B), at 0$^o$ to enforce a linear state of the source, followed by a quarter-wave plate, PSG QWP (Newport 10RP04-24), on a rotating stage (Thorlabs K10CR1/M) to control the PSG ellipticity through the PSG QWP angle $\Theta_{PSG}$. The samples were put on a glass slide with index matching oil to avoid scattering due to surface roughness and minimize lensing effects due to the sample geometry. The exit light is collected by an objective (depending on the measurement, Olympus UPLFLN4XP-2 x4/NA 0.13 or Olympus UPLFLN10XP-2 x20/NA 0.3). Polarization analysis is performed between the objective and the microscope frame tube lens to have parallel beams and to avoid angular artifacts on the polarization elements. The polarization states to analyze the sample (denoted PSA) are arranged in a symmetrical way to those of PSG with a quarter wave plate, PSA QWP (Newport 10RP04-24), on a rotating stage (Thorlabs K10CR1/M) to orient the later at $\Theta_{PSA}$, followed by a linear polarizer at $0^o$. The sensor, a CCD camera of 12bits 1920x1200 pixels with 5.86 $\mu$m square pitch (IDS U3-3060CP-M-GL), is placed at the image plane of the tube lens to collect the resulting intensity images, after the microscope built-in dielectric mirror at 45$^o$.


\section*{Data avalibility}
The raw data along with the scripts used to produce the figures in this paper are shared on the following doi:10.3929/ethz-b-000719385, https://www.research-collection.ethz.ch/handle/20.500.11850/719385.

\section*{Acknowledgements}
We thank Manfred Fiebig and Morgan Trassin for the fruitful discussions.

Preliminary diffraction experiments were performed on the BM01 at the ESRF. Diffraction data was measured on beamline I19 at Diamond Radiation Source.

AS and YK thank the funding from Swiss National Science Foundation PCEFP2\_203658 and PZ00P2\_180035. We acknowledge the use of LLMs ChatGPT and Claude for text editing and proofreading.

AB and ML received funding from the European Research Council Starting Grant MUMOTT (ERC-2020-SrG 949301), funded by the European Union. Views and options expressed are however those of the author(s) only and do not necessary reflect those of the European Union or the European Research Council Executive Agency. Neither the European Union nor the granting authority can be held responsible for them.

\section*{Author contributions statement}

A.S. and Y.K. conceived the experiments, Y.K., T.W., J.B., A.S., D.C. and M.W. performed synchrotron diffraction experiments, Y.K., A.B. and M.L. performed Mueller microscopy experiments, S.G. and I.B. performed high resolution diffraction experiments, Y.K., A.S. and A.B. performed data analysis. Y.K. and A.S. wrote the manuscript. All authors reviewed the manuscript.

\section*{Additional information}

The authors declare no competing interests. 

\newpage

\renewcommand{\thefigure}{S\arabic{figure}}
\setcounter{figure}{0} 

\renewcommand{\thetable}{S\arabic{table}}
\setcounter{table}{0} 

\section*{Supplementary Information}

\subsubsection*{Average structure and symmetry analysis}
\begin{figure*}[!ht]
\centering
\includegraphics[width=0.6\linewidth]{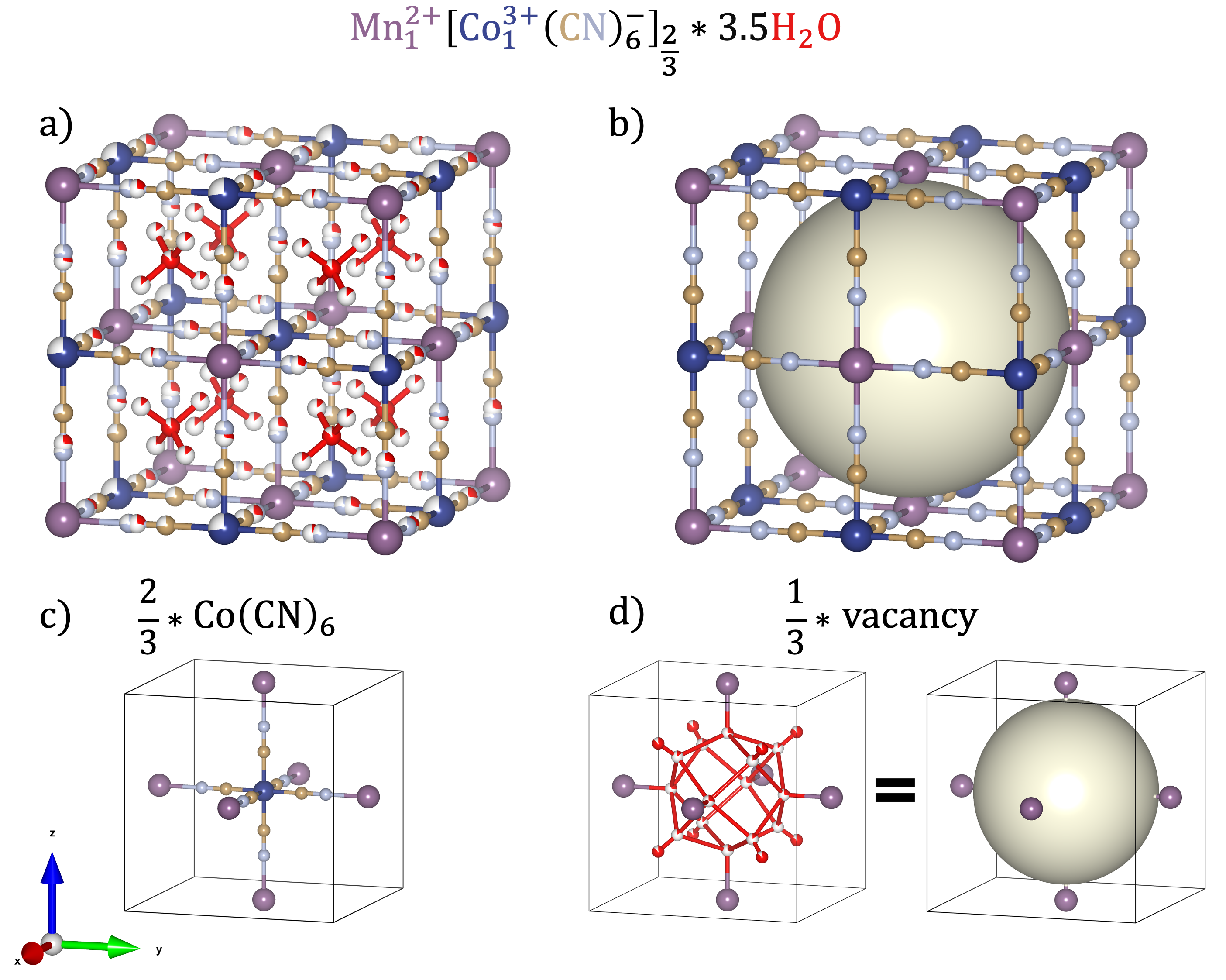}
\caption{a) Average structure of Mn[Co]-PBA, with space group $Fm\bar 3m$. Mn and Co occupy alternate vertices in the FCC lattice and are connected with CN linkers (c). 1/3 of Co(CN)$_6$ positions are vacant. The vacancy, represented by a beige sphere in the simplified representation (b), in the actual structure is filled with 14 water molecules, represented by red circles (d).}
\label{fig:av_str}
\end{figure*}

\begin{figure*}[!ht]
\centering
\includegraphics[width=\linewidth]{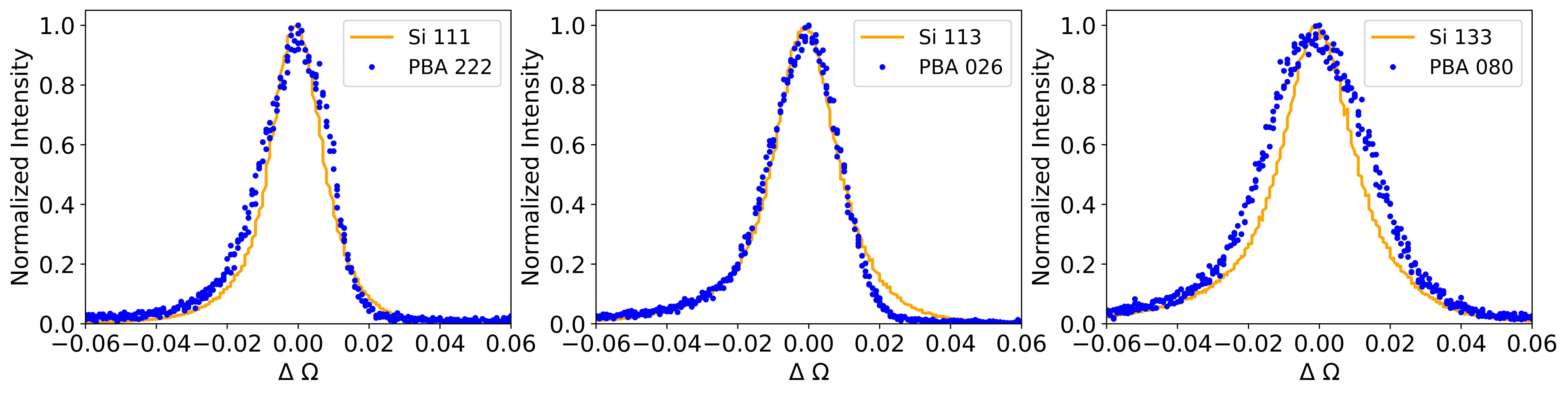}
\caption{Results of high-resolution X-ray diffraction measurement. The panels show the rocking curves for 222, 026, and 080 reflections of Mn[Co]-PBA crystal. Rocking curves show no splitting and the FWHM width of each measured curve is below 0.02$^{o}$, comparable to the widths of rocking curves of silicon crystal at similar angles. The intensity is integrated inside the detector and represented as a function of rocking angle ($\Delta$ $\Omega$).}
\label{fig:peak_splitting}
\end{figure*}

Mn[Co]-PBA crystals are known to adopt cubic average structure, represented in Figure \ref{fig:av_str}. The deviations from cubic symmetry due to strain or formation of ferroelastic domains can appear in peak splitting. High-resolution reciprocal space mapping is sensitive to such splitting. 
We conducted high-resolution reciprocal space mapping on 18 Bragg peaks. Three rocking curves of the measured peaks 222, 026, and 080 are displayed in Figure \ref{fig:peak_splitting}. All the measured peaks are sharp, we observed no peak splitting for all measured peaks. The width of the rocking curves is close to the instrument resolution and to the peak width of a silicon crystal. Thus, we see no indication of a reduction in symmetry in the average structure due to strain domains.

We further analyzed the average structure in Mn[Co]-PBAs by collecting single-crystal Bragg diffraction data from an untwinned portion of the crystal. 
Collected Bragg peaks were integrated and scaled using the XDS software. For this, we have chosen tetragonal symmetry, to ensure that the scaling procedure does not artificially increase the symmetry of the dataset. The resulting integrated peaks were left unmerged. The output file is \textit{scaf8\_unmerged.hkl}. This file was then converted by xdsconv into $res\_XSCALED.hkl$. The files can be found in the folder \textit{reconstruction/xds\_run0\_I4mmm}.

\begin{figure}[!ht]
\centering
\includegraphics[width=0.8\linewidth]{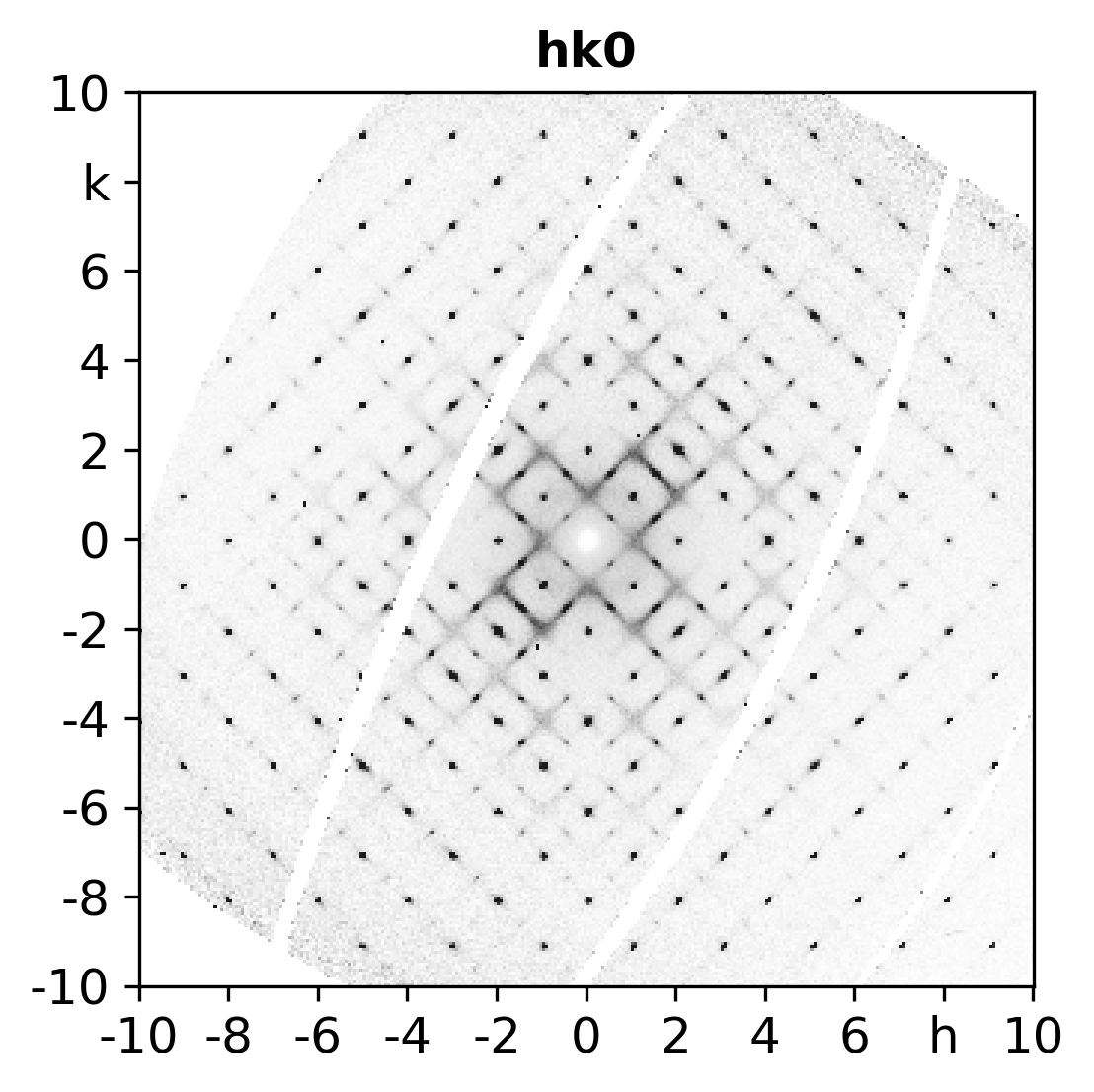}
\caption{Reconstructed hk0 plane of 3D dataset in tetragonal setting, with 4-fold rotation aligned with 00l direction.}
\label{fig:projection}
\end{figure}

The use of XDS software is not standard in the refinement of inorganic structures, but we have chosen it since it allowed us to perform Bragg peak integration and the reconstruction of reciprocal space using the same orientation matrix. This way we have ensured that the orientation of the tetragonal axis is correctly aligned along $\mathbf{c}$ axis, as can be seen on Figure \ref{fig:projection} or in three-dimensional reconstructed data (folder \textit{reconstruction/reconstruct}).

Next, we have used Jana2020 \cite{petvrivcek2023jana2020} to transform the tetragonal reflection file to the cubic setting (file \textit{03\_transformed\_to\_cubic.hkl}. The unit cell parameters in two given space groups, $Fm\bar 3m$ and $I4/mmm$, are provided in table \ref{table:unit_cell_params}. 

Next, we refined tetragonal ($I4/mmm$) and cubic ($Fm\bar 3m$) structures in ShelxL. The structure refinement in cubic symmetry resulted in the structure with parameters provided in table \ref{table:str_cub}. We have used anisotropic atomic displacement parameters for Co, Mn, C, N and oxygen atoms which belong to the structural water, and isotropic atomic displacement parameters for the zeolitic water.

The refinements were conducted with the following constraints: 
\begin{itemize}
    \item the occupancies of Co, C, and N atoms were set equal; 
    \item the occupancies of structural water atoms O2 and O3 were set to 1-occ(Co);
    \item the atomic displacement parameters (ADP) of N and O atoms at shared positions were set equal: ADP(N1)=ADP(O2) and ADP(N2)=ADP(O3).
\end{itemize}

\begin{figure}[!ht]
\centering
\includegraphics[width=0.6\linewidth]{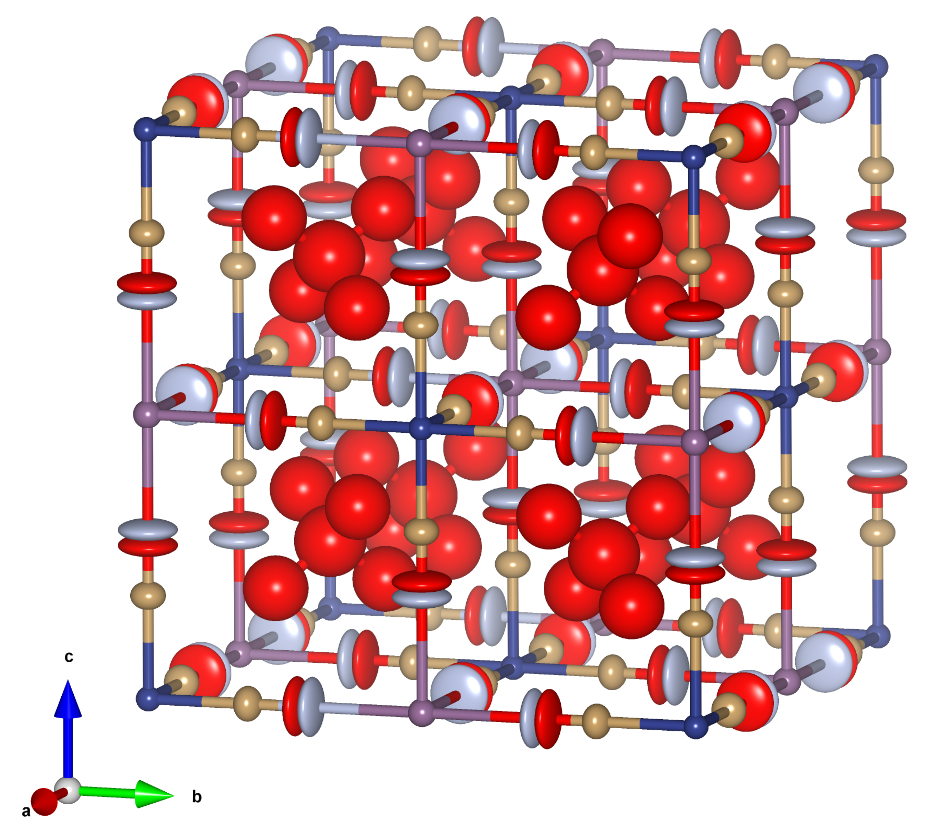}
\caption{Result of structure refinement of Mn[Co]-PBA in cubic $Fm\bar 3m$ space group.}
\label{fig:cubic_cubic}
\end{figure}

\begin{figure*}[!ht]
\centering
\includegraphics[width=0.7\linewidth]{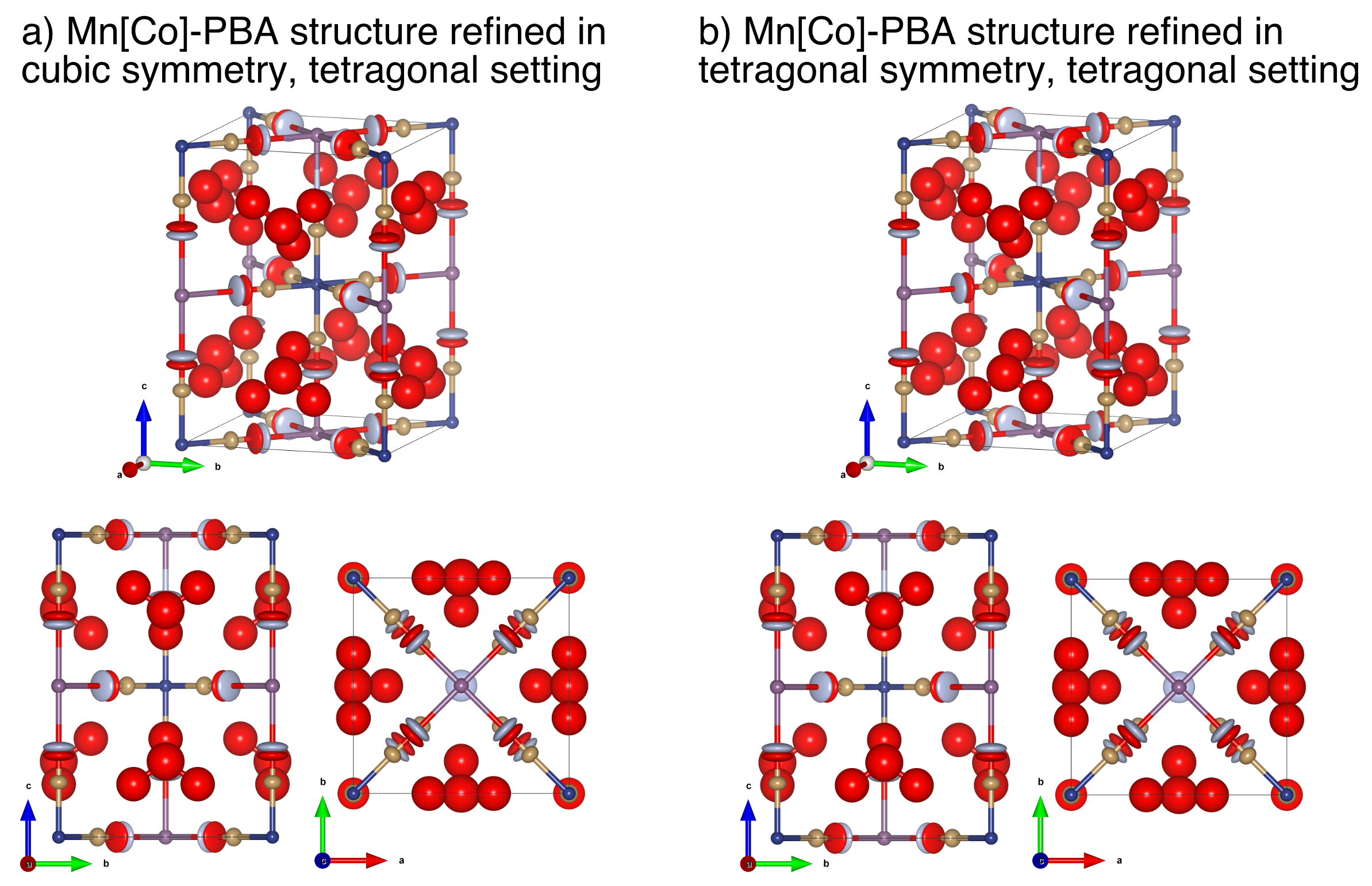}
\caption{The resulting refined structures in cubic $Fm\bar 3m$ and tetragonal $I4/mmm$ symmetry compared in the tetragonal setting.}
\label{fig:cubic_vs_tetragonal}
\end{figure*}

\begin{figure*}[!ht]
\centering
\includegraphics[width=0.5\linewidth]{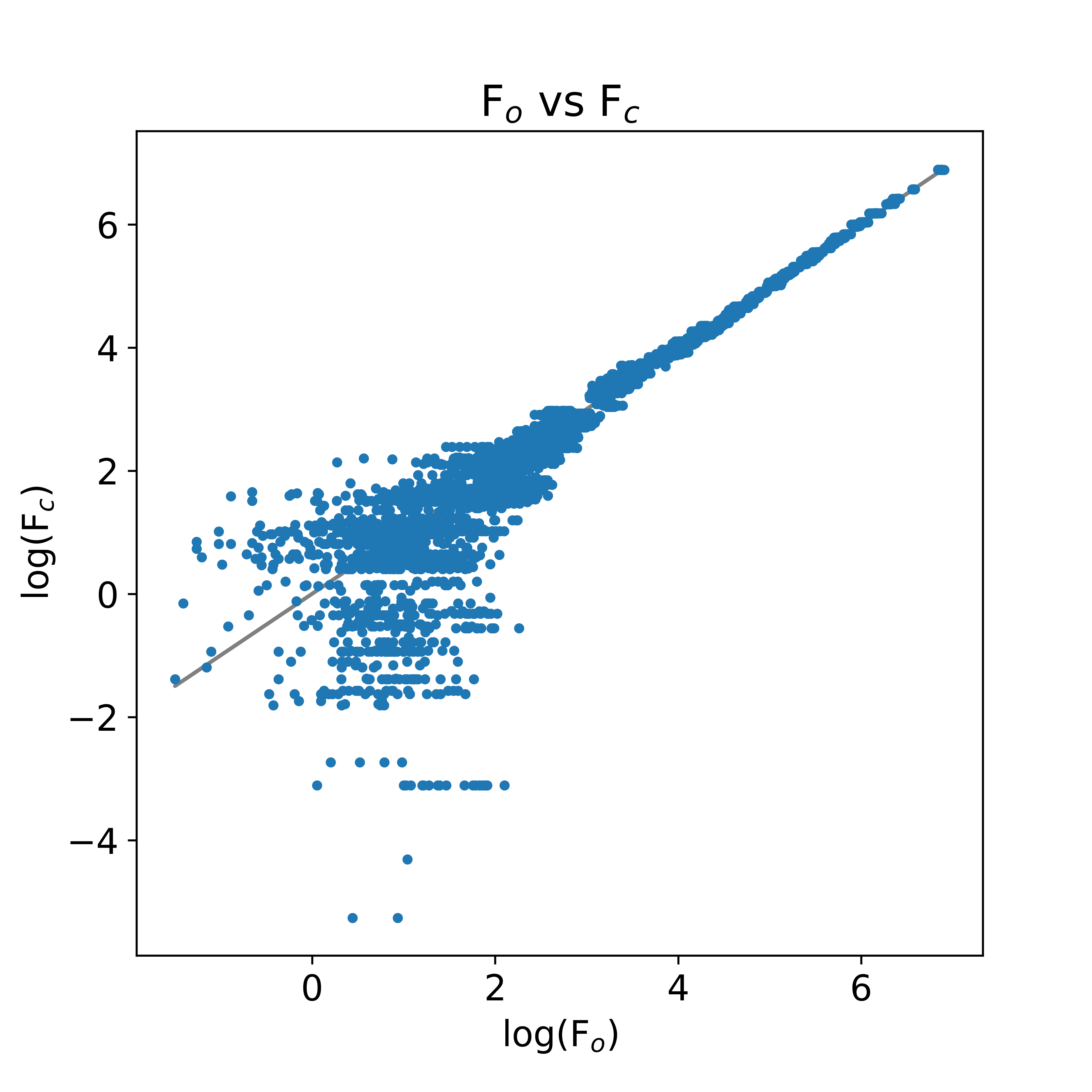}
\caption{F$_{o}$ vs F$_{c}$ plot for the tetragonal structure refinement against 8126 peaks.}
\label{fig:fobs_calc}
\end{figure*}

During refinement, we made sure that ShelxL \cite{sheldrick2008short} did not merge the data (option MERG 0) such that R-factors of cubic and tetragonal structures are calculated from the identical set of reflections. We also did not use the reweighing scheme (WGHT 0) because weighting schemes use calculated intensities which are slightly different in cubic and tetragonal structure, which would mean that the weights in the two models will be different. The extinction parameter was refined but showed a relatively low value of 0.003 (tetragonal) and 0.001 (cubic).
Figure \ref{fig:cubic_cubic} displays the refined structure of Mn[Co]-PBA crystal in cubic symmetry represented in the usual cubic setting. To simplify the comparison of refined structures in two symmetries, we represent the refined cubic structure in the tetragonal setting (Figure \ref{fig:cubic_vs_tetragonal}a). The refined structure in tetragonal symmetry is represented in Figure \ref{fig:cubic_vs_tetragonal}b. The details of the refined structures in cubic and tetragonal symmetries are provided in the tables \ref{table:str_cub_tetr} and \ref{table:str_tetr}, respectively. The final parameters of the refinements are listed in table \ref{table:refinem_params} with F$_{obs}$ vs F$_{calc}$ plot for tetragonal refinement displayed in figure \ref{fig:fobs_calc}.

\begin{table*}[!ht]
\centering
\caption{Unit cell constants in two given spacegroups}
\begin{tabular}{ | c | c | c | c | c | c | c |}
  \hline
SG  & a, \AA & b, \AA & c, \AA & $\alpha$, $^o$ & $\beta$, $^o$ & $\gamma$, $^o$ \\ \hline            
    $Fm\bar 3m$ & 10.411(8) & 10.411(8) & 10.411(8) & 90  & 90  & 90  \\ \hline   
      $I4/mmm$ & 7.362(4) & 7.362(4) & 10.411(8) & 90  & 90 & 90  \\ \hline

\end{tabular}
\label{table:unit_cell_params}
\end{table*}

\begin{table*}[!ht]
\centering
\caption{Refined fractional coordinates and atomic displacement parameters at 300K according to cubic $Fm\bar3m$ space-group setting.}
\begin{tabular}{| c | c | c | c | c | c | c | c |}
  \hline
  Atom & Occupancy & x & y & z & U11 & U22 & U33 \\ 
  \hline            
  Co1 & 0.716(3) & 0 & 0 & 0 & \multicolumn{3}{c|}{0.0221(4)} \\ 
  \hline   
  Mn1 & 1 & 0.5 & 0 & 0 & \multicolumn{3}{c|}{0.0282(3)} \\ 
  \hline   
  C1  & 0.716(3) & 0.8186(4) & 0 & 0 & 0.0286(16) & 0.0440(13) & 0.0440(13) \\ 
  \hline   
  N1  & 0.716(3) & 0.7030(5) & 0 & 0 & 0.016(3) &  0.125(3) & 0.125(3)\\ 
  \hline   
  O1  & 0.8599 & 0.75 & 0.75 & 0.75 & \multicolumn{3}{c|}{0.181(16)} \\ 
  \hline   
  O2  & 0.284(3) & 0.7303(14) & 0 & 0 & 0.016(3) & 0.125(3) & 0.125(3) \\ 
  \hline   
  O3  & 0.14 & 0.8252(18) & 0.6748(18) & 0.6748(18) & \multicolumn{3}{c|}{0.149(11)} \\ 
  \hline   
\end{tabular}
\label{table:str_cub}
\end{table*}

\begin{table*}[!ht]
\centering
\caption{Final refinement, cubic in tetragonal setting.}
\begin{tabular}{| c | c | c | c | c | c | c | c |}
  \hline
  Atom & Occupancy & x & y & z & U11, U22 & U12 & U33 \\ 
  \hline            
  Co1 & 0.716(3) & 0 & 0 & 0 & \multicolumn{3}{c|}{0.02210(7)} \\ 
  \hline   
  Mn1 & 1 & 0.5 &0.5 & 0 & \multicolumn{3}{c|}{0.02820(7)} \\ 
  \hline   
  C1  & 0.716(3) & 0.8186(2) & 0.1814(2) & 0 & 0.0363(6) & 0.0389(3) & 0.0440(6)  \\ 
  \hline   
  C2  & 0.716(3) & 0.5 & 0.5 & 0.3186(2) & 0.0440(4) & 0.0389(4) & 0.0286(10)\\ 
  \hline   
  N1  & 0.716(3) & 0.7030(3) & 0.297(3) & 0 & 0.0705(11) & 0.0887(7) & 0.1250(16)\\ 
  \hline  
  N2  & 0.716(3) & 0.5  & 0.5  & 0.2030(3) & 0.1250(11) & 0.0887(7) & 0.0160(15) \\ 
  \hline  
  O1  & 0.8599 & 0  & 0.5  & 0.25 & \multicolumn{3}{c|}{0.181(5)} \\ 
  \hline   
  O2  & 0.284(3) & 0.7303(9) & 0.2697(9) & 0 & 0.0705(11) & 0.0887(7) & 0.1250(16)  \\ 
  \hline   
  O3  & 0.284(3) & 0.5  & 0.5 &  0.2303(9) & 0.1250(11) & 0.0887(7) & 0.0160(15)  \\ 
  \hline   
  O4  & 0.14 & 0.150(2) & 0.5 & 0.3252(16) & \multicolumn{3}{c|}{0.149(7)} \\ 
  \hline   
\end{tabular}
\label{table:str_cub_tetr}
\end{table*}

\begin{table*}[!ht]
\centering
\caption{Final refinement, tetragonal. Refined fractional coordinates and atomic displacement parameters at 300K according to $I4/mmm$ space-group setting.}
\begin{tabular}{| c | c | c | c | c | c | c | c |}
  \hline
  Atom & Occupancy & x & y & z & U11, U22 & U12 & U33 \\ 
  \hline            
  Co1 & 0.716(2) & 0 & 0 & 0 & 0.0224(3) & 0.0221(2) & 0.0214(3) \\ 
  \hline   
  Mn1 & 1 & 0.5 & 0.5 & 0 & 0.0284(2) & 0.0282(2) & 0.0279(3) \\ 
  \hline   
  C1  & 0.716(2) & 0.8186(3) & 0.1814(3) & 0 & 0.0370(9) & 0.0391(6) & 0.0432(16)\\ 
  \hline   
  C2  & 0.716(2) & 0.5 & 0.5 & 0.3186(5) & 0.0438(14) & 0.0385(9) & 0.0278(17) \\ 
  \hline   
  N1  & 0.716(2) & 0.7030(4) & 0.297(4) & 0 & 0.0705(19) & 0.0881(16) & 0.123(3) \\ 
  \hline  
  N2  & 0.716(2) & 0.5  & 0.5  & 0.2030(5) & 0.126(3) & 0.089(2) & 0.016(4) \\ 
  \hline  
  O1  & 0.86 & 0  & 0.5  & 0.25 & \multicolumn{3}{c|}{0.181(10)} \\ 
  \hline   
  O2  & 0.284(2) & 0.7306(11) & 0.2694(11) & 0 & 0.0705(19) & 0.0881(16) & 0.123(3) \\ 
  \hline   
  O3  & 0.284(2) & 0.5  & 0.5 &  0.2293(18) & 0.126(3) & 0.089(2) & 0.016(4) \\ 
  \hline   
  O4  & 0.14 & 0.151(3) & 0.5 & 0.325(2) & \multicolumn{3}{c|}{0.149(7)} \\ 
  \hline   
\end{tabular}
\label{table:str_tetr}
\end{table*}

\begin{table}[!ht]
\centering
\caption{Final parameters of refinement in cubic $Fm\bar 3m$ and tetragonal $I4/mmm$ space-group settings.}
\begin{threeparttable}
\begin{tabular}{| c | c | c |}
  \hline
  Parameter & $Fm\bar3m$  & $I4/mmm$ \\ 
  \hline  
  $R_{int}^{a}$,  (\%) & 3.28 & 3.16 \\  
  \hline  
  GOF$^{b}$(obs) & 2.252 & 2.234 \\  
  \hline  
  R1$^{c}$(obs),  (\%) & 3.64 & 3.59 \\  
  \hline  
  wR2$^{d}$(obs),  (\%) & 12.45 & 12.34 \\  
  \hline  
  GOF$^{b}$(all) & 2.252 & 2.234 \\  
  \hline  
  R1$^{c}$(all),  (\%)  & 3.9 & 3.84 \\ 
  \hline  
  wR2$^{d}$(all),  (\%) & 12.87 & 12.76 \\  
  \hline  
  Number of parameters refined, p & 15 & 27 \\  
  \hline
  Number of reflections, n & 8126 & 8126 \\  
  \hline
\end{tabular}
\begin{tablenotes}
  \footnotesize
  \item[a] Reliability factor of measured dataset $R_{int}$ is defined as $R_{int}=\sum|F^2_o-F^2_o(mean)|/\sum|F^2_o|$, where $F^2_{o}$ means the experimental measured scattering intensity; $F^2_{c}$ is calculated intensity; $F_{o}$ and $F_{c}$ are observed and calculated structure factors, respectively;
  \item[b] Goodness of fit $GOF=S=\{\sum[w(F^2_o-F^2_c)^2]/(n-p) \}^{1/2}$, where n is the number of reflections; p is the total number of parameters refined; $w=1/[\sigma^2(F_o^2)+(aP)^2 + bP]$; $P$ is $[2F_c^2+Max(F_o^2,0)]/3$.
  \item[c] Reliability factor $R1=\sum||F_o|-|F_c||/\sum|F_o|$.
  \item[d] Weighted residual factor $wR2=\{\sum [w(F^2_o-F^2_c)^2]/\sum[w(F^2_o)^2]\}^{1/2}$.
\end{tablenotes}
\end{threeparttable}
\label{table:refinem_params}
\end{table}

\subsubsection*{3D-$\Delta$PDF refinement}

We analyze the local structure of Mn[Co]-PBA crystals using Yell software\cite{simonov2014yell}, which relies on the Three Dimensional Difference Pair Distribution function (3D-$\Delta$PDF) approach \cite{weber2012three}. 3D-$\Delta$PDF provides information about features of the real structure that are not represented in the average structure, describing these features in terms of interatomic pair correlations. Using this method, we run the refinements of the pair correlations for building blocks separated at different lattice vectors versus the experimental diffuse scattering. 

\begin{figure}[!ht]
\centering
\includegraphics[width=0.9\linewidth]{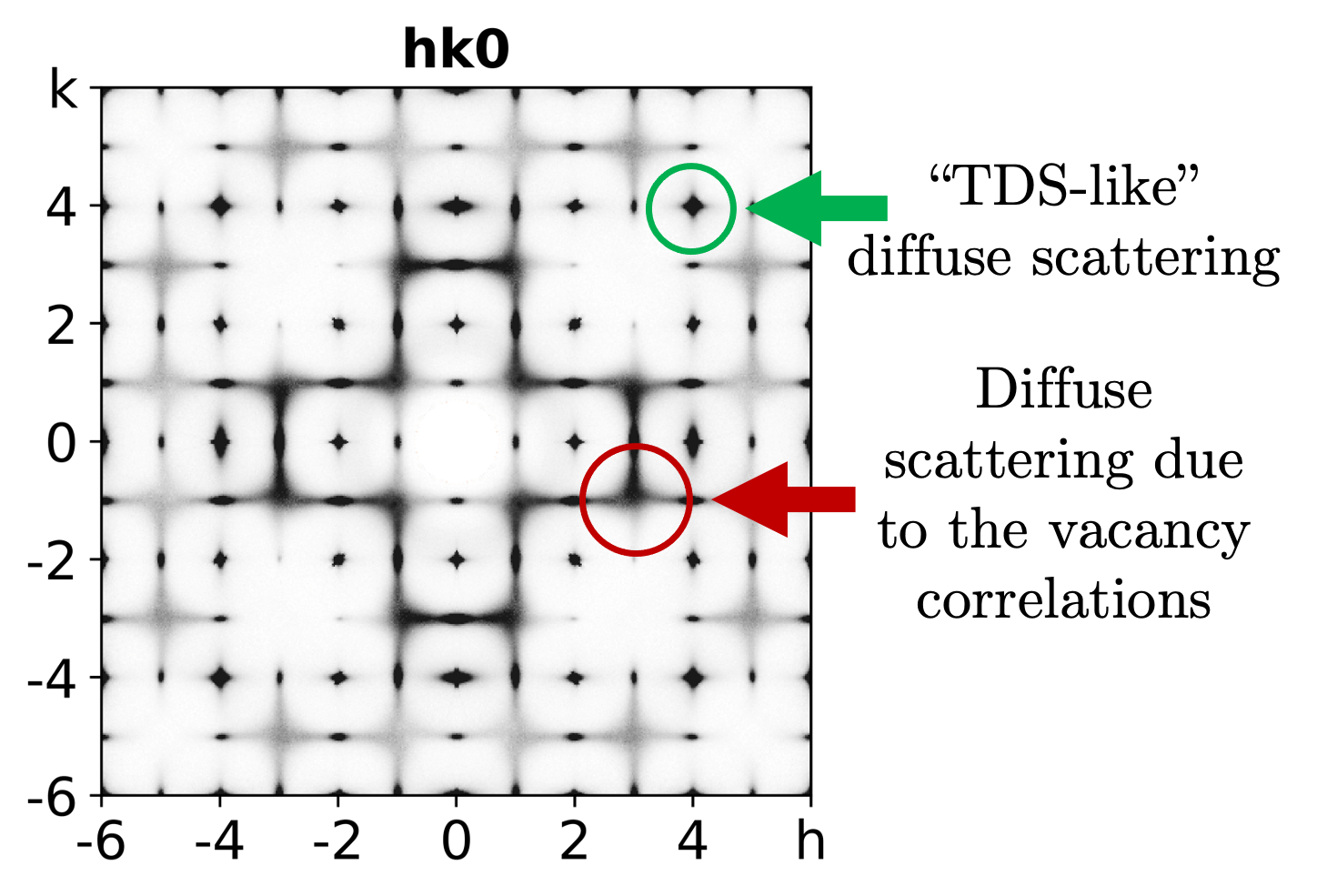}
\caption{Reconstructed (hk0) plane of diffuse scattering dataset. This dataset contains three main components: Bragg peaks at even h,k,l; broad diffuse scattering around Bragg peaks referred to as "thermal diffuse scattering-like" (green arrow); and crosses of diffuse scattering at the borders of the Brillouin zone (red arrow), which arises due to non-random Co(CN)$_6$ distribution.}
\label{fig:diffuse_w_arrows}
\end{figure}

Experimentally measured diffuse scattering from Mn[Co]-PBAs consists of two parts: 
\begin{itemize}
    \item broad diffuse scattering around Bragg peaks often referred to as "thermal diffuse scattering-like" (TDS-like); in this crystal it arises mainly due to correlated static displacements in the framework (figure \ref{fig:diffuse_w_arrows}, green arrow); 
    \item crosses of diffuse scattering centered at the borders of the Brillouin zone (figure \ref{fig:diffuse_w_arrows}, red arrow). This diffuse scattering originates from the correlations in vacancy occupancy. 
\end{itemize}
    
Thus, the experimental diffuse scattering pattern can be described directly in terms of pair correlations of substitutional and displacive type. Substitutional correlation (SC) means that the occupancy of one site depends on the occupancy of the other site. It is represented by the probability of finding both sites occupied simultaneously. Displacive correlation (DC) between two sites is defined as a covariance of their translational modes.

The 3D-$\Delta$PDF model of Mn[Co]-PBA is refined against the experimental three-dimensional reconstructed diffuse scattering with a rather low resolution, 200x200x200 pixels. This low resolution is required to ensure the reasonable execution time of the program and the operating memory required. Both strongly depend on the number of pixels in the experimental diffuse scattering map, the number of interatomic pairs to be refined, and the number of parameters in the refinement \cite{simonov2014yell}.

\begin{figure*}[!ht]
\centering

\begin{subfigure}{\textwidth}
\includegraphics[width=\linewidth]{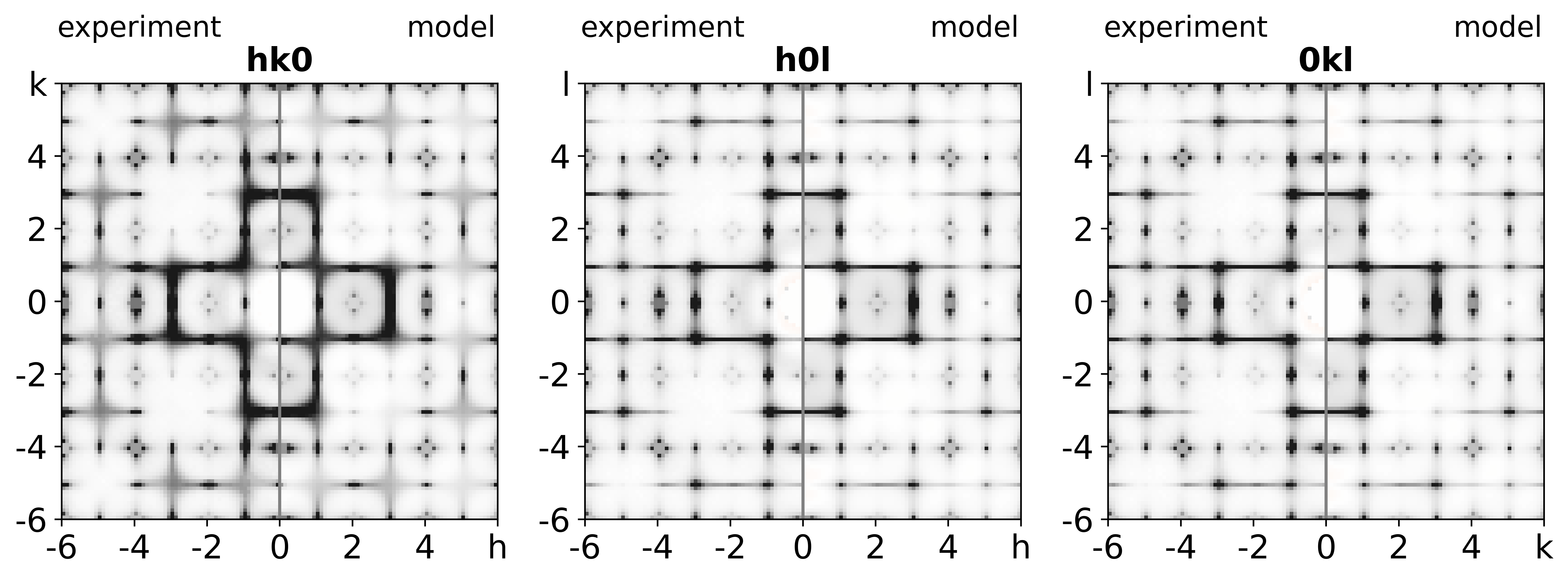}
\includegraphics[width=\linewidth]{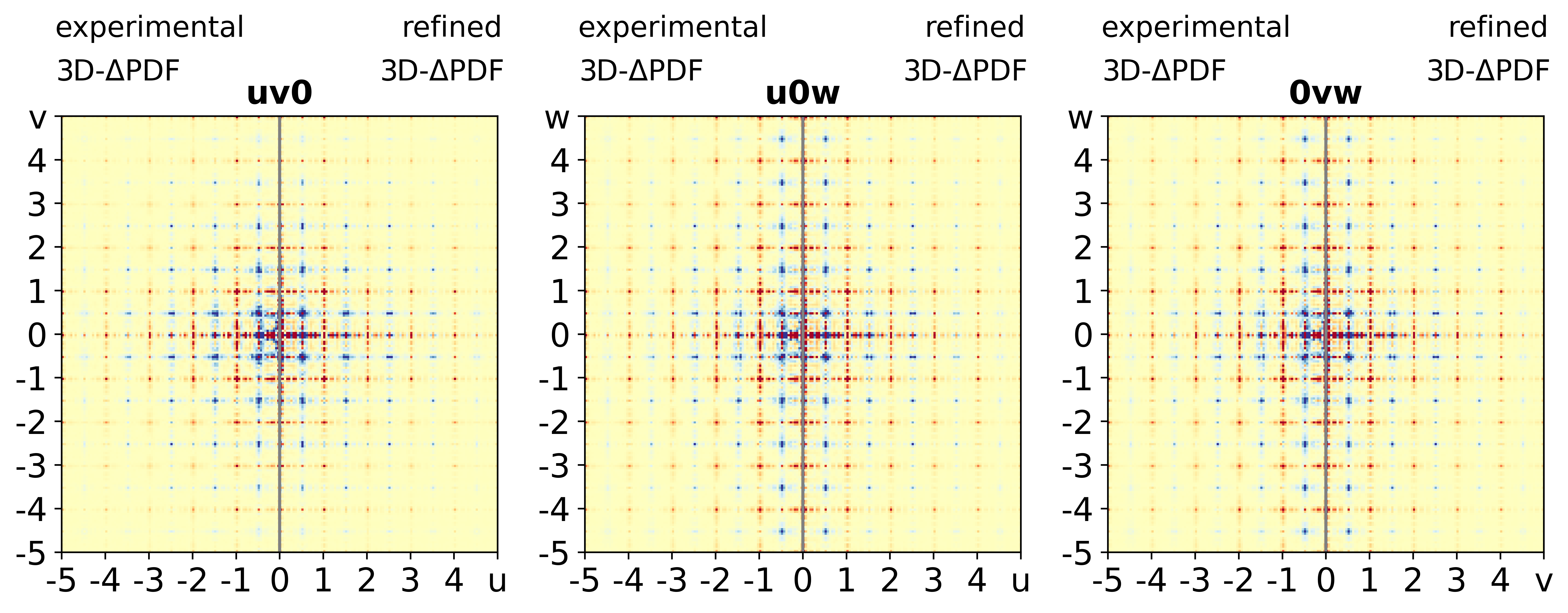}
\end{subfigure}

\caption{Comparison of the experimental data with the refined model in reciprocal (a) and 3D-$\Delta$PDF space (b). The model includes substitutional and displacive pair correlations for all neighbor pairs up to the 5th unit cell.}
\label{fig:diffuse_modelled_PDF}
\end{figure*}

\begin{figure*}[!ht]
\centering

\begin{subfigure}{\textwidth}
\includegraphics[width=\linewidth]{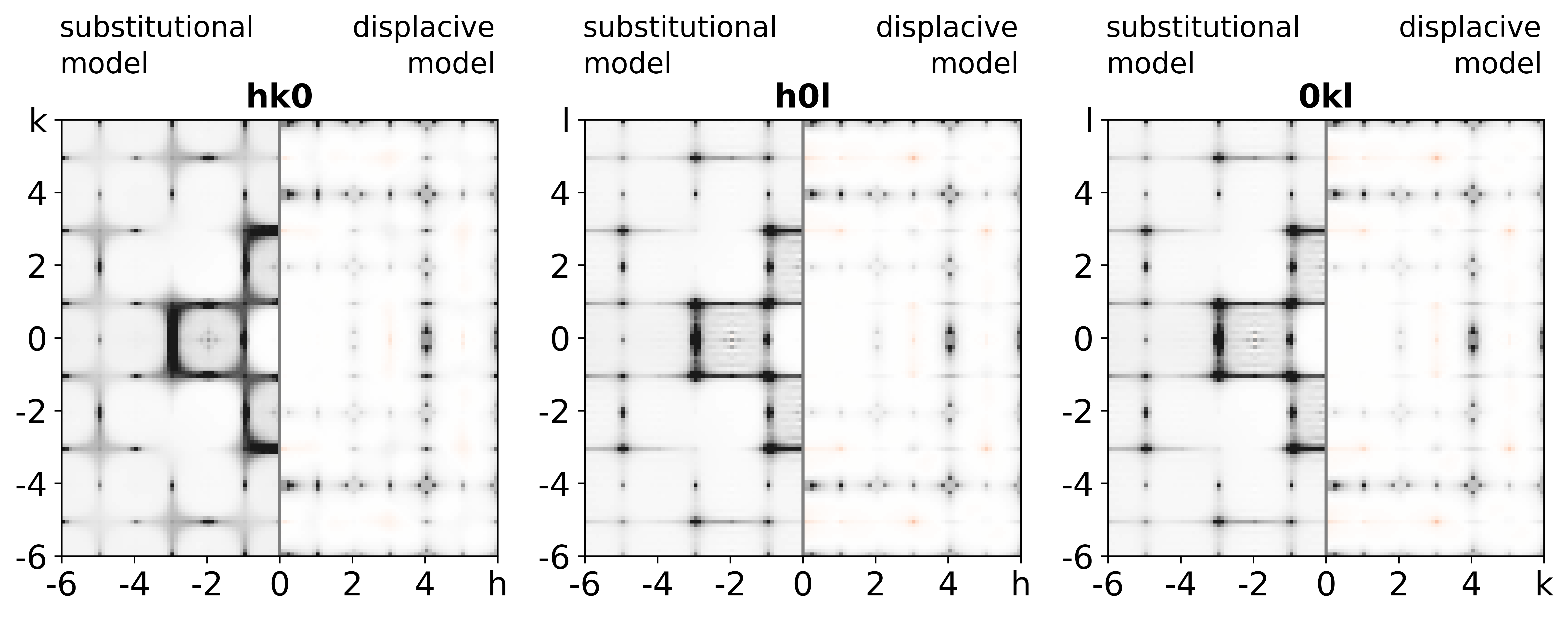}
\includegraphics[width=\linewidth]{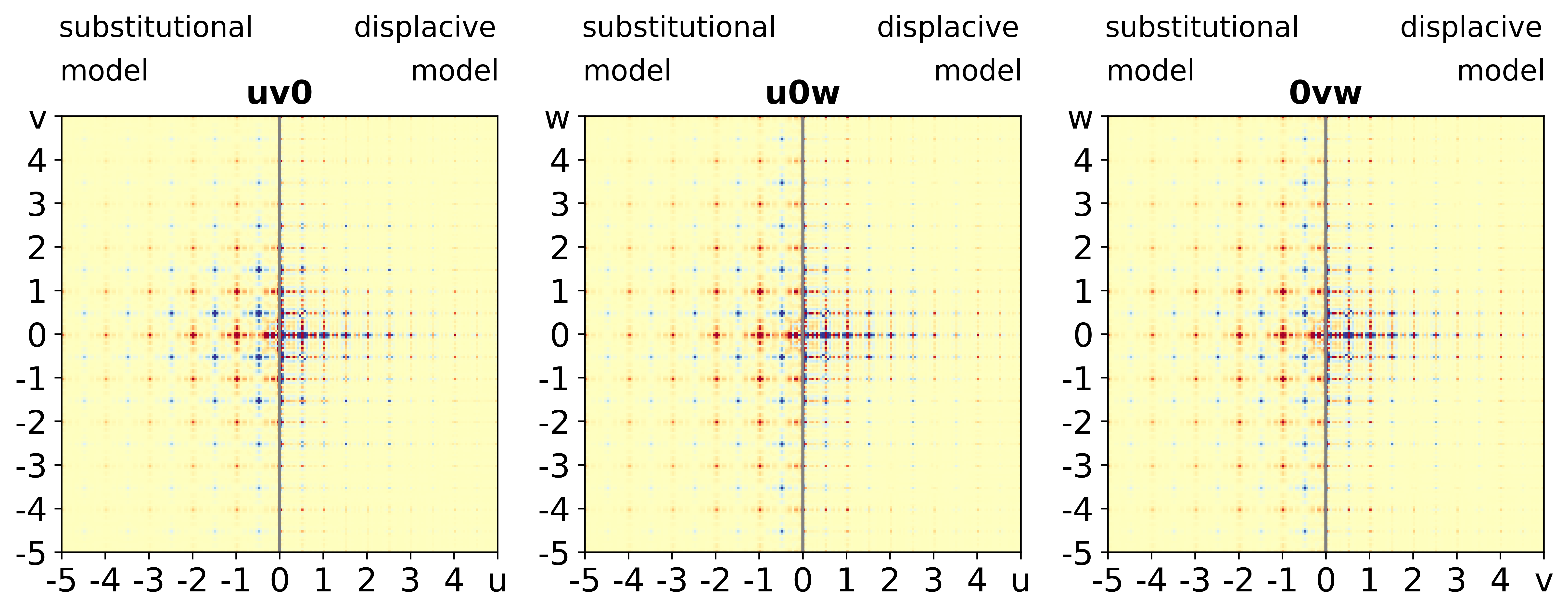}
\end{subfigure}

\caption{Models of substitutions and displacive pair correlations represented in reciprocal space (a) and 3D-$\Delta$PDF space (b). The 3D-$\Delta$PDF refinement of the experimental dataset was performed using the program Yell. Two types of correlations were refined: substitutional and displacive, for all pairs of neighbors in the model. Substitutional correlations describe the diffuse crosses at the Brillouin zone borders, while displacive correlations reproduce the diffuse scattering around Bragg peaks.}
\label{fig:models_separated}
\end{figure*}

TDS signal is broad, it partially overlaps with the diffuse due to vacancy correlations. This creates a substantial challenge in refining the local structure model of the substitutional type of disorder. We mitigate this problem by refining two types of disorder, substitutions and displacive simultaneously. To run the local structure refinement, we prepare the dataset by removing the Bragg peaks from the reconstructed experimental diffuse scattering dataset, represented in figure \ref{fig:diffuse}. We remove the Bragg peaks with a punch and fill procedure using the software Meerkat \cite{simonov-meerkat}. The Bragg peaks in the given dataset are sharp, so we chose the punched and filled area to be a sphere of a 2-pixel radius. Next, using yell software we run 3D-$\Delta$PDF refinement for two kinds of correlations, substitutions type, and displacive type, for all the pair of building blocks within a distance of 5*5*5 unit cells radius. This simultaneous refinement of two types of correlations reproduces well the experimental data, figure \ref{fig:diffuse_modelled_PDF}, and allows disentangling of the portion of the diffuse signal that arises due to non-random disorder in vacancy occupancies from the displacive disorder, figure \ref{fig:models_separated}. 

The refined model contains enough parameters to refine not only the experimental diffuse scattering but also fit the artifacts created by the punch and fill procedure, figure \ref{fig:diffuse_modelled_PDF}. However, these artifacts don't affect the substitutional correlations in a model and are only contained in the parameters of displacive correlations, which we don't use in the further analysis of this paper. It becomes evident that the substitutional correlations are unaffected by these artifacts if we separate the refined model into two portions: one with the substitutional correlations and one with the displacive correlations alone, figure \ref{fig:models_separated}.  The refined substitutional correlations reproduce the experimental crosses of diffuse scattering and capture the loss of cubic symmetry. The refined models are provided in the supplementary file.

\begin{table}[!ht]
\centering
\caption{Result of 3D-$\Delta$PDF refinement: Substitutional pair correlations for the neighbors, separated by the vector (x,y,z). See the supplementary file "correlation\_matrix.npz" for the complete list of pair correlations.}
\begin{tabular}{| c | c | c | c |}
  \hline
  x & y  & z & refined pair probability, \%\\ 
  \hline  
  0 & 0 & 0 & 0.806 \\  
  \hline  
  
  0.5 & 0.5& 0& -0.189 \\  
  \hline 
  0& 0.5& 0.5& -0.132\\  
  \hline  
  
  1& 0& 0& 0.214 \\  
  \hline  
  0& 0& 1& 0.273 \\  
  \hline
  
  1& 0.5& 0.5& 0.032 \\  
  \hline
  0.5& 0.5& 1& -0.041 \\  
  \hline
  
  1& 1& 0& 0.091 \\  
  \hline  
  1 & 0 & 1 & 0.108 \\  
  \hline  
  
  1& 1& 1& 0.024 \\  
  \hline  

  1.5& 0.5& 0& -0.107 \\  
  \hline
  1.5& 0& 0.5& -0.072 \\  
  \hline
  
  1& 1.5& 0.5& -0.005 \\  
  \hline
  1.5& 0.5& 1& -0.049 \\  
  \hline
  1& 0.5& 1.5& 0.010 \\  
  \hline

  1.5& 1.5& 0& -0.055 \\  
  \hline  
  1.5& 0& 1.5& -0.042 \\  
  \hline

  2& 0& 0& 0.104 \\  
  \hline
  0& 0& 2& 0.186 \\  
  \hline

\end{tabular}

\label{table:pair_correlations}
\end{table}

\subsubsection*{Monte Carlo Modelling}

Monte Carlo simulations were carried out using custom-written code, provided. 
Each configuration is represented by a 16x16x16 supercell. Configurations were initialized with a random distribution of vacancies, such that one-third of the sites were vacant. 
We represent Monte Carlo energy of the generated structure by the Hamiltonian (Eq. \ref{eq:Hamiltonian}):

\begin{align}
&    H = \dfrac{1}{2} \sum_{r\in\{M\}} {e}_{r} \Bigg[J_{1} \sum_{r'\in{\frac{1}{2}\langle110\rangle}} {e}_{r+ r'}
    + J_{2,x,y} \sum_{r'\in{\{[100], [010]\}}} {e}_{r\pm r'}+ \nonumber \\
&    + J_{2,z} \sum_{r'\in{\{[001]\}}} {e}_{r\pm r'}\Bigg],
    \label{eq:Hamiltonian}
\end{align}

where the sum is taken over all M sites at position r with the neighboring sites at $r+ r'$ with states $e_{r+ r'}=0$ (vacant) or $e_{r+ r'}=1$ (present). 
The interactions in this Hamiltonian are set to: 
\begin{itemize}
    \item the nearest neighbor interaction $J_{1}=0.605$
    \item the second neighbors interactions $J_{2,x}=J_{2,y}=-0.0135$, and $J_{2,z}=-0.11$
\end{itemize}
For the given interaction parameters, we generate 5,000 structures. The Monte Carlo step involves a swap of one vacancy ($e_{r+ r'}=0$) with one hexacyanocobaltate cluster $e_{r+ r'}=1$, both chosen at random. The swap is accepted if the energy of the system is reduced, if it increases, the swap is accepted with the probability $p=\exp\Big[\dfrac{E1-E2}{T}\Big]$, where E1 - is the energy of the system before the swap, E2 - is energy after the swap and $T$ is Monte Carlo temperature, set to 1.3. The configurations were equilibrated for 500 epochs (one epoch corresponds to the number of steps required to visit each site twice on average) with 20 epochs steps in between each produced structure.

\begin{figure*}[!ht]
\centering
\includegraphics[width=\linewidth]{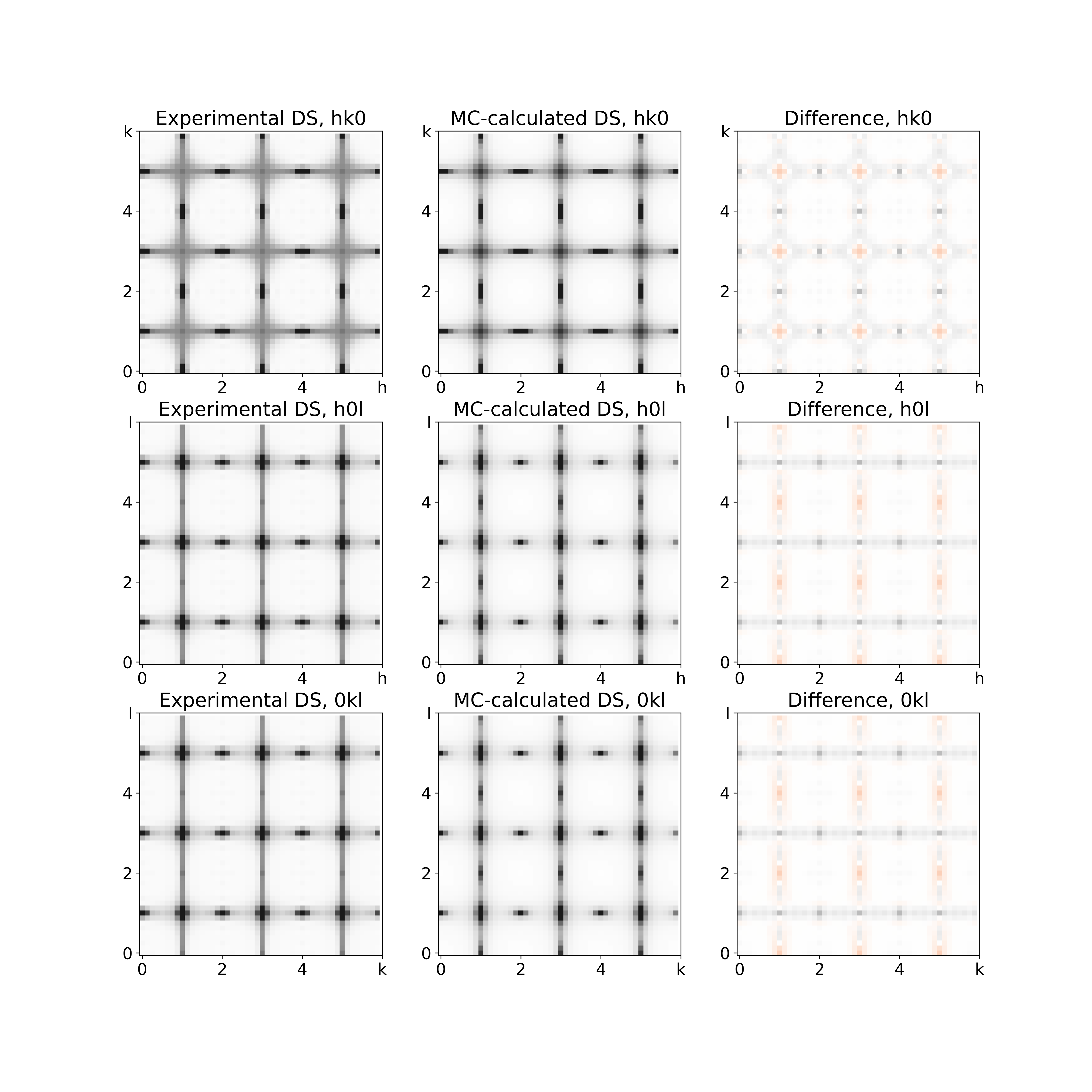}
\caption{Comparison of the experimental diffuse scattering (Experimental DS) with diffuse scattering calculated from the structures generated by Monte Carlo (MC-calculated DS). We represent only the structure factor portion of diffuse scattering, the form-factor portion is omitted.  This Monte Carlo Model relies upon interaction parameters between the nearest (J$_{1}$) and second nearest (J$_{2}$) neighbors. The first term, J$_{1}$, describes vacancy interactions between nearest neighbors along <110> directions, with J$_{1}$=0.605 for [110], [101], and [011] directions. The second term, J$_{2}$, describes vacancy interactions along <100> directions, split into J$_{2,x}$, J$_{2,y}$, and J$_{2,z}$ to reduce symmetry from cubic to tetragonal, with J$_{2,x}$=J$_{2,y}$=-0.0135 for the second nearest neighbors along [100] and [010] directions, and J$_{2,z}$=-0.11 for the second nearest neighbor along [001] direction. The Monte Carlo temperature T was set to 1.3. The R-factor for this model is 21.9\%.}
\label{fig:MC}
\end{figure*}

Figure \ref{fig:MC} shows the agreement of experimental diffuse scattering with diffuse scattering calculated from the structures generated by Monte Carlo. In this work, we aimed to get a reasonable agreement between the model and the experimental data, using the minimum number of interaction parameters, to maintain model simplicity.

\subsubsection*{Mueller Polarimetry}
\begin{figure*}[!ht]
\centering
\includegraphics[width=0.8\linewidth]{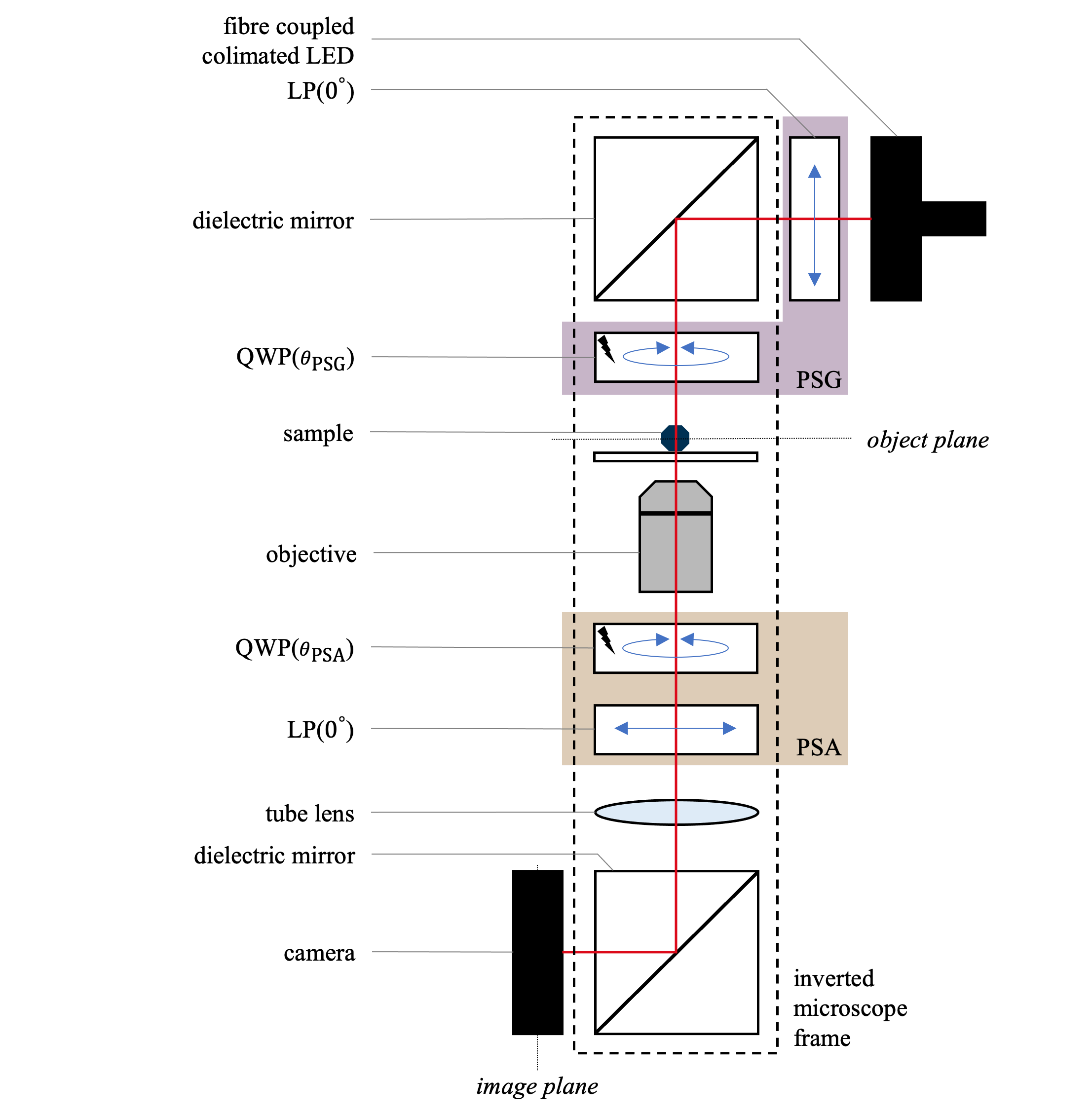}
\caption{Setup for Mueller Polarimetry.}
\label{fig:MP_setup}
\end{figure*}

The data were acquired on the setup in Figure \ref{fig:MP_setup}, using an exposure of 30ms with the x20/NA 0.3 objective and 8ms for the x4/NA 0.13 (this is subject to variation depending on the sample. See the sample descriptions for more precision). Due to the objective magnification, numerical aperture, and the sensor pixel pitch, the acquisition was performed at a resolution of 0.586 $\mu$m (respectively 1.465$\mu$m for the x4 objective).
For each field of view, we performed a polarimetric scan using the sixteen combinations of PSG [LP $0^{o}$, EL (0.5, $0^{o}$), RCP, LCP] and PSA [LP $0^{o}$, EL (0.5, $0^{o}$), RCP, LCP], recording the intensity image resulting for each [S1]. Note that EL (0.5, $0^{o}$) stand for an elliptical state of ellipticity +0.5, aligned at $0^{o}$, RCP is right circular polarisation, and LCP left circular polarisation. 
From the recorded intensity maps, we can extract the Mueller matrix of the samples at every pixel \cite{chipman2018polarized}, and deduce from the latter their generalized birefringence properties using the Lu $\&$ Chipman polar decomposition method \cite{lu1996interpretation}.

\end{document}